# Anisotropic critical-state model of type-II superconducting slabs

Yingxu Li[†] and Yuanwen Gao[*]

*Key Laboratory of Mechanics on Environment and Disaster in Western China, The Ministry of Education of China, Lanzhou, Gansu 730000, P. R. China*

*Department of Mechanics and Engineering Science, College of Civil Engineering and Mechanics, Lanzhou University, Lanzhou, Gansu 730000, P. R. China*

[*] Corresponding author: ywgao@lzu.edu.cn

**Abstract**: We introduce a critical-state model incorporating the anisotropy of flux-line pinning to analyze the critical states developing in an anisotropic biaxial superconducting slab exposed to a uniform perpendicular magnetic field and to two crossed in-plane magnetic fields which are applied successively. The theory is an extension of the anisotropic collective pinning theory developed by Mikitik and Brandt. The anisotropic flux-line pinning enters into the critical states by generating the angular dependence of the critical current density and by deviating the direction of the electric field from the current in the plane perpendicular to the vortex line. Comparing to the isotropic case, the anisotropic flux pinning strongly influences the magnetic response in the slab. We also apply the critical-state model to predict the underlying physical phenomenon of a field-cooled slab in a rotating magnetic field.

## 1. Introduction

The critical-state model in Type-II superconductors is first introduced by Bean [1, 2]. The component $J_\perp$ of the current density **J** perpendicular to the magnetic induction **B** is restricted to the threshold $J_{c\perp}$ for flux depinning, $J_\perp = J_{c\perp}$. The vortex moves whenever the driving Lorentz $f_L = J_\perp B$ exceeds the average pinning force $f_p = J_{c\perp} B$, thus causing a local flux-transport electric field [3]. As it only contains physics of flux pinning, Bean model solves the critical states in a symmetric superconductor exposed to an external magnetic field that is applied along the symmetry axis [1] [4, 5] [6] [7-9]. The Bean critical states also describe the most cases of practical interest, such as in superconducting magnets where the self-field is generally perpendicular to the local current flow [10] and in HTS ac power cables [11].

If **J** is not perpendicular to the magnetic field **H** ($\mathbf{B} = \mu_0 \mathbf{H}$ is a good approximation for high-$\kappa$ superconductors [12]), as expected in the samples with nonsymmetrical geometry or excited by the magnetic field that varies in both magnitude and direction [13-16], the vortex lines may tilt each other. To avoid flux-line cutting, there is a maximum gradient of the tilt angle $\alpha$ such that the component $J_\parallel$ of **J** along **H** is constrained to the threshold $J_{c\parallel}$. The fundamental physics is the local helical instability in a vortex with a sufficiently large $J_\parallel$ [17-19]. Flux-line cutting explains the first onset of a nonvanishing parallel component $E_\parallel$ of the electric field at $|J_\parallel| \geq J_{c\parallel}$ [20]. This is put into the generalized double-critical-state model (GDCSM) by Clem and the coauthors [21-23], accounting for both flux cutting and flux transport in the critical states. Assuming uncoupled flux transport and flux cutting, GDCSM sketches a rectangular of $J_c(J_\parallel, J_\perp)$ in the $J_\parallel - J_\perp$ plane. GDCSM successfully predicts the experimental magnetic response of high-temperature superconducting plates [24] and YBCO thin film [25] in which **B** is inclined with **J**. In practical applications, this theory allows to calculate the AC loss in power transmission cables with second generation HTS wires [26].

---

[†] Current address: School of Mechanics and Engineering, Southwest Jiaotong University, Chengdu, Sichuan 610000, People's Republic of China.

An extension of GDCSM is made by Brandt and Mikitik [27], who incorporate the coupling of flux depinning and cutting into the critical-state model. In their proposal, the rectangular $J_c(J_\parallel, J_\perp)$ changes into a quasi-ellipse. Romero-Salazar and Pérez-Rodríguez introduce an elliptic critical-state model [13, 28, 29] to explain the observations of a smooth angular dependence of $J_c$ in experiments [30, 31]. This model also accounts for the interdependent flux depinning and cutting, both of which simultaneously occur everywhere at the boundary of the critical-state ellipse except for $J_\perp = 0$ and $J_\parallel = 0$, accompanied by the appearance of $E_\perp$ and $E_\parallel$. The material law **E(J)** is assumed to follow $E_\perp/E = J_\perp/J_{c\perp}$ and $E_\parallel/E = J_\parallel/J_{c\parallel}$. The elliptic critical-state model is then extended to account for the general relations between $E$ and $J$ for $|J|$ just above $J_c$, $E_\perp = \rho_\perp J_\perp$ and $E_\parallel = \rho_\parallel J_\parallel$ where $\rho_\perp$ and $\rho_\parallel$ are the nonlinear effective resistivities [12]. A generalized framework describing the critical state of type-II superconductors is the variational statements [32-36], which incorporates appropriate conductivity laws into the differential Maxwell equations. The variational method allows the extension of the double-critical-state model to three-dimensional configurations. This method exposes the critical states in superconducting slabs [34, 36], wires [37, 38] and practical HTS cables [39] subject to a wide range of external magnetic fields and transport currents. A numerical method using the vector potential and commercial software is developed to solve the critical state in bulks under crossed fields, and then extended to coils in two and three dimensions [40].

In light of the anisotropic extensions of Bean's model as well as of the variational statements by Badía-López-Ruiz, we extend the critical-state theory to account for anisotropic flux-line pinning in anisotropic biaxial type-II superconductors. We focus on the *T*-critical states where only flux pinning occurs [21], $J_\perp = J_{c\perp}$ and $J_\parallel < J_{c\parallel}$, in terms of $J_{c\parallel} \gg J_{c\perp}$. Whereas the Bean critical states and the double critical states are only the limiting cases, the *T*-critical states occur in the most cases [21, 22] [41] [42] [43, 44]. The anisotropy of flux-line pinning entering in *T*-critical states has been preliminarily considered in thin disk-shaped superconductors [45] and superconducting elliptic films [46] exposed to a perpendicular magnetic field, and in long strips and rectangular platelets of high-$T_c$ superconductors with in-plane and out-of-plane pinning anisotropy [41], and also in infinite slabs and strips exposed to arbitrarily oriented magnetic field [27]. Here, we give a quantitative analysis of the critical states in anisotropic superconducting slabs, on the strength of a recent microscopic characterization [47, 48] of the anisotropic pinning in the framework of the collective pinning theory [49-52].

We start with the analyses of the critical states in isotropic superconductors in Section 2. The analyses of anisotropic case in Section 3 is highly relevant to the isotropic one. We then incorporate the anisotropic flux-line pinning in the critical-state model and analyze the magnetic response of an anisotropic biaxial type-II slab. In Section 4, we use the critical-state model to analyze the magnetic response in a slab subject to an in-plane rotating magnetic field. At last, we summarize the main results of the paper. Throughout the paper, we use $\mathbf{H}$, $\mathbf{J}$ and $\mathbf{E}$ to denote the field vectors. $\hat{\mathbf{H}}$ and $H$ represent the unit vector and the magnitude of $\mathbf{H}$. Also we have $\hat{\mathbf{J}}$ and $J$ for $\mathbf{J}$ as well as $\hat{\mathbf{E}}$ and $E$ for $\mathbf{E}$.

## 2. Isotropic critical-state model

*2.1 Material conductivity law*

We assume there are randomly distributed point pinning centers in isotropic superconductors and $J_{c\perp}$ is independent with $B$. The periodic nucleation of quantized vortices at the surface and their self-annihilation when penetrating in generate the first appearance of a steady-state electric field $\mathbf{E}$, $\mathbf{E} = \mathbf{B} \times \mathbf{v}_v$ with $\mathbf{v}_v$ the velocity of the vortex [3]. When the Lorentz force $\mathbf{J}_\perp \times \mathbf{B}$ locally exceeds the pinning force, the vortex starts to move at the driving force direction assuming there is no Hall angle [52], $\mathbf{v}_v \parallel [\mathbf{J}_\perp \times \mathbf{B}]$. Thus, the electric field $\mathbf{E}$ is expressed as

$$\mathbf{E} = E\hat{\mathbf{E}} = E\hat{\mathbf{J}}_\perp. \quad (1)$$

The magnitude of the electric field, $E$, is expressed via a voltage-current law accounting for the flux-flow loss. In the microscopic point of view, motion of the vortex is described by the balance of the Lorentz force against the viscous drag force [53-55], and the resistivity law for flux transport is $E_\perp = \rho_\perp J_\perp$ where $\rho_\perp$ is the flux-flow resistivity. From the macroscopic phenomenological perspective, the static critical state due to flux pinning is correlated to flux transport regime through any model law $E(J)$ which has a sharp rising at $J = J_c$. $E(J)$ may take the form of the power law, $E = E_0 (|J_\perp|/J_{c\perp})^n$, while in the Bean model it appears as: $E = 0$ for $J_\perp < J_{c\perp}$, $E = E_0$ for $J_\perp = \text{sgn}(E_\perp)J_{c\perp}$ and $E \to \infty$ for $J_\perp > J_{c\perp}$. Here $E_0$ is a finite value of $E$.

*2.2 Analyses of longitudinal transport problem*

We now study the longitudinal transport problem (LTP) in an infinite slab, with the lateral dimensions $L$ much larger than the thickness $a$ of the sample. To proceed with our analysis, we introduce a Cartesian coordinate system ($O-xyz$) with its $xy$ plane coinciding with the middle plane of the sample, Fig. 1. All dimension lengths of interest in the sample far exceeds the flux-line spacing $a_0$, and we are allowed to treat the sample as a homogenous medium. We assume that all the macroscopic electromagnetic quantities ($\mathbf{J}$, $\mathbf{E}$ and $\mathbf{H}$) vary in the slab $xy$ plane by a characteristic scale which is much larger than the thickness $a$, i.e. one may neglect the derivatives of $\mathbf{J}$, $\mathbf{E}$ and $\mathbf{H}$ with respect to $x$ and $y$. This assumption is reasonable for many infinite thin samples (rectangular plate, trip and circular disk), and the scale of the variation in the $x-y$ plane is indeed of the same order of the lateral dimensions $L$ except for a small region near the flux front [41].

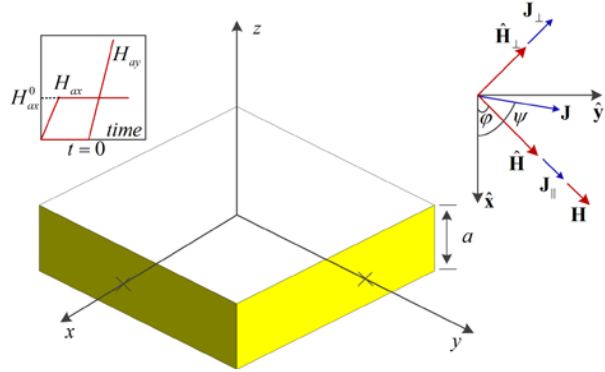

**Figure 1.** Schematic of the longitudinal transport problem in an infinite slab. Insert: Vectors used in this problem. The coordinate base vectors are the in-plane $\hat{\mathbf{x}}$, $\hat{\mathbf{y}}$ and out-of-plane $\hat{\mathbf{z}}$. The orthogonal base vectors are $\hat{\mathbf{H}}$ and $\hat{\mathbf{H}}_\perp$. The magnetic field $\mathbf{H} = H\hat{\mathbf{H}}$ with an angle $\varphi$ relative to $\hat{\mathbf{x}}$. The current density $\mathbf{J} = J_\perp \hat{\mathbf{H}}_\perp + J_\parallel \hat{\mathbf{H}}$ at an angle $\psi$.

Under the magneto-quasi-stationary (MQS, slow and uniform sweep rates of the external magnetic sources) regime, the time-dependent Maxwell equations read as

$$\nabla \times \mathbf{H} = \mathbf{J}, \quad \partial_t \mathbf{B} = -\nabla \times \mathbf{E}, \tag{2}$$

with integrability conditions

$$\nabla \cdot \mathbf{B} = 0, \quad \nabla \cdot \mathbf{J} = 0. \tag{3}$$

Here, the displacement current densities $\partial_t \mathbf{D}$ representing charge separation and recombination vanish in a first-order treatment. When $H \gg H_{c1}$ at the surface, $\mathbf{B} = \mu_0 \mathbf{H}$ is a good approximation for high-$\kappa$ superconductors [12].

*2.2.1 T-state equations*

Let us introduce a pair of orthogonal unit vectors $\hat{\mathbf{H}}$ and $\hat{\mathbf{H}}_\perp$ which direct along and perpendicular to the magnetic field $\mathbf{H} = H\hat{\mathbf{H}}$, Fig. 1. We resolve the current density $\mathbf{J}$ along $\hat{\mathbf{H}}$ and $\hat{\mathbf{H}}_\perp$,

$$\mathbf{J} = J\hat{\mathbf{J}} = \mathbf{J}_\perp + \mathbf{J}_\parallel = J_\perp \hat{\mathbf{H}}_\perp + J_\parallel \hat{\mathbf{H}}, \tag{4}$$

with $J_\perp = \mathbf{J} \cdot \hat{\mathbf{H}}_\perp$ and $J_\parallel = \mathbf{J} \cdot \hat{\mathbf{H}}$. Since $\mathbf{H}$ and $\mathbf{J}$ lie in the slab plain, the current density in the plane $\perp \mathbf{H}$ equals to the component $\mathbf{J}_\perp$ in the slab plane. Then, using the conductivity law (1), the electric field is

$$\mathbf{E} = E_\perp \hat{\mathbf{H}}_\perp. \tag{5}$$

The external excitation $\mathbf{H}_a$ is parallel to the $xy$ plane, $\mathbf{H}_a = H_{ax}\hat{\mathbf{x}} + H_{ay}\hat{\mathbf{y}}$ at $z = \pm 0.5a$. Recalling the conservation law (3), one has $H_z' = J_z' = 0$ such that $H_z = J_z = 0$. $\hat{\mathbf{H}}$ and $\hat{\mathbf{J}}$ in $O-xyz$ are

$$\hat{\mathbf{H}} = \hat{\mathbf{x}}\cos\varphi + \hat{\mathbf{y}}\sin\varphi, \quad \hat{\mathbf{J}} = \hat{\mathbf{x}}\cos\psi + \hat{\mathbf{y}}\sin\psi, \tag{6}$$

where $\varphi$ and $\psi$ signify the angles of $\mathbf{H}$ and $\mathbf{J}$ with respect to $x$ axis. Using orthogonality relations $\hat{\mathbf{H}} \cdot \hat{\mathbf{H}}_\perp = 0$ and $\hat{\mathbf{H}} \times \hat{\mathbf{H}}_\perp = \hat{\mathbf{z}}$, we have $\hat{\mathbf{H}}_\perp$ in $O-xyz$,

$$\hat{\mathbf{H}}_\perp = -\hat{\mathbf{x}}\sin\varphi + \hat{\mathbf{y}}\cos\varphi. \tag{7}$$

The projections $J_\parallel$ and $J_\perp$ of the current $\mathbf{J}$ on the directions $\hat{\mathbf{H}}$ and $\hat{\mathbf{H}}_\perp$ now can be expressed as

$$J_\parallel = (\hat{\mathbf{J}} \cdot \hat{\mathbf{H}})J = J\cos(\psi-\varphi), \quad J_\perp = (\hat{\mathbf{J}} \cdot \hat{\mathbf{H}}_\perp)J = J\sin(\psi-\varphi). \tag{8}$$

Thus, in $O-xyz$, Ampere's law has two projections on $\hat{\mathbf{H}}$ and $\hat{\mathbf{H}}_\perp$, $-H\varphi' = J_\parallel$, $H' = J_\perp$, where $H' \equiv \partial H/\partial z$ and $\varphi' \equiv \partial\varphi/\partial z$. They are consistent with the appropriate formulas in [21, 41]. Similarly, Faraday's law yields two equations $E_\perp \varphi' = \mu_0 H \dot{\varphi}$ and $E_\perp' = \mu_0 \dot{H}$, where $\dot{\varphi} \equiv \partial\varphi/\partial t$ and $\dot{H} \equiv \partial H/\partial t$. At $T$-critical states, we have $J_\perp = (E_\perp/E)J_{c\perp} = \text{sgn}(E_\perp)J_{c\perp}$ and $J_\parallel = J_\perp/\tan(\psi-\varphi)$ according to Eqs. (5) and (8). We are now ready to set down the $T$-state equations appropriate for LTP in an isotropic slab,

$$\text{Isotropic } T\text{-state equation in LTP} \begin{cases} \text{I}: -H\varphi' = J_\parallel = \text{sgn}(E_\perp)J_{c\perp}/\tan(\psi-\varphi), \\ \text{II}: H' = J_\perp = \text{sgn}(E_\perp)J_{c\perp}, \\ \text{III}: E_\perp \varphi' = \mu_0 H \dot{\varphi}, \\ \text{IV}: E_\perp' = \mu_0 \dot{H}. \end{cases} \tag{9}$$

*2.2.2 C-state and CT-state equations*

Compared to $T$-critical state, there are two different aspects in $C$-critical states (flux-cutting critical states): first $\mathbf{E}$ is directed along $\hat{\mathbf{H}}$ rather than $\hat{\mathbf{H}}_\perp$, $\mathbf{E} = E_\parallel \hat{\mathbf{H}}$, and second the critical-state restriction is $J_\parallel = J_{c\parallel}$ rather that $J_\perp = J_{c\perp}$. If the sample is in the $CT$-states where both flux transport and cutting occur, we have $|J_\perp| = J_{c\perp}$ and $|J_\parallel| = J_{c\parallel}$ while the direction of the electric field $\mathbf{E}$ is implicit. We are now ready to summary the equations for $C$- and $CT$-critical states:

$$\text{Isotropic } C\text{-state equation in LTP} \begin{cases} -H\varphi' = J_\parallel = \text{sgn}(E_\parallel)J_{c\parallel}, \\ H' = J_\perp = \text{sgn}(E_\parallel)J_{c\parallel}\tan(\psi-\varphi), \\ E_\parallel \varphi' = \mu_0 \dot{H}, \\ E_\parallel' = -\mu_0 H \dot{\varphi}, \end{cases} \tag{10}$$

$$\text{Isotropic CT-state equation in LTP} \begin{cases} -H\varphi' = J_{\parallel} = \text{sgn}(E_{\parallel})J_{c\parallel}, \\ H' = J_{\perp} = \text{sgn}(E_{\perp})J_{c\perp}, \\ E'\sin(\psi_E - \varphi) + E\psi_E'\cos(\psi_E - \varphi) = \mu_0 \dot{H} \text{ or } E_{\perp}' + E_{\parallel}\varphi' = \mu_0 \dot{H}, \\ E'\cos(\psi_E - \varphi) - E\psi_E'\sin(\psi_E - \varphi) = -\mu_0 H\dot{\varphi} \text{ or } E_{\parallel}' - E_{\perp}\varphi' = -\mu_0 H\dot{\varphi}. \end{cases} \quad (11)$$

If all electromagnetic quantities are time-independent, the *CT*-state equation is reduced to the critical-state equations in GDCSM [23]. Note, however, these equations (9), (10) and (11) are specialized in infinite slab, and the generalization is needed for other geometries.

*2.2.3 T-, C- and CT- critical states*

Let us now find the fundamental physics in the interdependence between *T*-, *C*- and *CT*- critical state based on GDCSM, Fig. 2, and the critical-state equations (9), (10) and (11). Due to the existence of external magnetic sources $H_a(t)$, a finite electric field **E** is induced which causes the onset of a current **J** inside the O zone, $J < J_c$. Initially, **J** is not large enough to break the critical current threshold $J_c$, and at this stage it is served as a current flow in an idea conductor which follows $\partial_t \mathbf{J} \propto \mathbf{E}$. In the O zone, neither flux transport nor flux-line cutting occurs. Then, **J** arrives at the boundary of the O zone. Here we assume $J_{c\parallel} \ll J_{c\perp}$ So, **J** first reaches $J_{\perp} = \pm J_{c\perp}$ and the *T*-critical state attains. The selection of $+$ or $-$ for $J_{\perp}$ coincides with the relative direction between **E** and $\hat{\mathbf{H}}_{\perp}$: $+$ stands for $\mathbf{E} \cdot \hat{\mathbf{H}}_{\perp} = \text{sgn}(E_{\perp}) > 0$ ($T_+$ state) and $-$ for $\mathbf{E} \cdot \hat{\mathbf{H}}_{\perp} = \text{sgn}(E_{\perp}) < 0$ ($T_-$ state). These are incorporated in the *T*-state equation (9), in which the electric field **E** is parallel to $\hat{\mathbf{H}}_{\perp}$, and the critical-state condition is imposed by $J_{\perp} = \text{sgn}(E_{\perp})J_{c\perp}$ and $J_{\parallel} = \text{sgn}(E_{\perp})J_{c\perp}\tan^{-1}(\psi - \varphi)$.

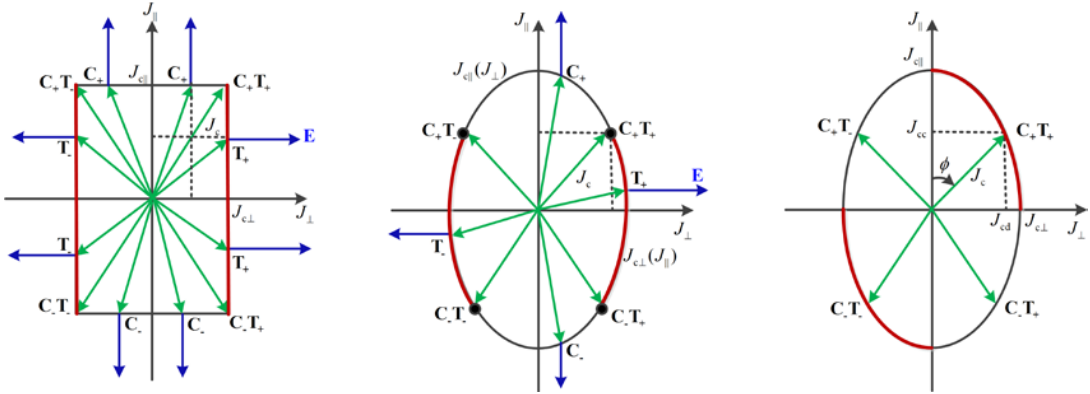

**Figure 2.** Isotropic critical-state models sketched in the $J_{\parallel} - J_{\perp}$ plane. Left: GDCSM [21-23] with irrelevant $J_{c\parallel}$ and $J_{c\perp}$. Center: BM critical-state model [27] with interdependent $J_{c\parallel}$ and $J_{c\perp}$. Right: The extended elliptical critical model [12].

For the first-order sweep, $\dot{H}$ is a constant, and according to the *T*-state equation IV, $E_{\perp}$ exhibit a linear gradient distribution independent with time. There is a linear distribution of the electric field independent with time for one-direction sweep, but a time-dependent nonlinear distribution for two-direction sweep. The latter can be taken as a quasilinear one under MQS regime. On the other hand, $\dot{H} > 0$ leads to $E_\perp > 0$ since $E_\perp' > 0$ and $E_\perp = 0$ at the flux front. Therefore, $\text{sgn}(E_\perp)$ is determined from the critical-state evolution, and the equation II gives the magnetic field profile. From this point one may recover the 1D Bean model, upon which the electric field enters the magnetic process via its direction, $H' = \pm J_{c\perp}$. In fact, the *T*-state equation differs significantly from the Bean model: first, the *T*-state equation involves two-dimensional physical quantities, for instance **H** characterized by its magnitude $H$ and azimuth $\varphi$; second, the double-critical-state-model is incorporated here; third, **E** enters into the *T*-state equation by not only the $\text{sgn}(E_\perp)$ but $E_\perp$ which affects the evolution of $\varphi$. To be specific, the *T*-state equation III dictates a variation of $\varphi$ with time and location, causing a simultaneous evolution in angle $(\psi - \varphi)$ ruled by the equation I. **J** changes progressively with its endpoint constrained to $J_\perp = \pm J_{c\perp}$, until it reaches the vertices of the critical-state rectangle.

In the *CT*-critical state, **J** is restricted to the vertices with $J = \sqrt{J_{c\perp}^2 + J_{c\|}^2}$ and $\tan(\psi - \varphi) = \pm J_{c\|}/J_{c\perp}$. The magnetic field are described with the *CT*-state equation (11): the equation II gives its magnitude $H$, and the equation I determines its orientation $\varphi$. The electric field **E** is a little more implicit but still determinate so long as its magnitude $E$ and direction $\psi_E$ are coupled in the *CT*-state equations III and IV. $\psi_E$ may change with time and location under the evolution of $H$, $\varphi$ and $E$. In case of $\psi_E - \varphi = \pm 0.5\pi$, **E** is changed into $\pm J_\|$ axis, and thus the sample enters into *C*-critical states. Following this idea, we have the reversible cycle paths for evolution of the critical states, $T_+ \rightleftarrows C_+T_+ \rightleftarrows C_+ \rightleftarrows C_+T_- \rightleftarrows T_- \rightleftarrows C_-T_- \rightleftarrows C_- \rightleftarrows C_-T_+ \rightleftarrows T_+$.

A generalization [27] of GDCSM by Brandt and Mikitik (BM critical-state model) incorporates the coupling effect of flux-line transport and cutting in the creep activation barrier, $U_{\text{barr}} = U_{\text{barr}}(J_\perp, J_\|)$. The two regimes occur simultaneously, which means the functions $J_{c\perp} = J_{c\perp}(J_\|)$ and $J_{c\|} = J_{c\|}(J_\perp)$ hold true at $J_\perp = J_{c\perp}$ and $J_\| = J_{c\|}$. In the $J_\perp - J_\|$ plane, Fig. 2, $J_{c\perp} = J_{c\perp}(J_\|)$ and $J_{c\|} = J_{c\|}(J_\perp)$ sketch two pairs of curved segments; the segments close to the $J_\perp$ axis representing the *T*-states and the segments close to the $J_\|$ axis representing the *C*-states. These curves cross one another at four isolated points corresponding to the CT states. The directions of the electric fields for different critical states are specified in accordance with those in GDCSM. The evolution of the critical states follows the same rules as in the rectangular critical state, except for $J_{c\perp} = J_{c\perp}(J_\|)$ and $J_{c\|} = J_{c\|}(J_\perp)$.

*2.2.4 Calculations*

Introducing the normalizations $h = 2H/(J_{c\perp}a)$, $e_\perp = 4E_\perp t_0/(\mu_0 J_{c\perp}a^2)$, $j = J/J_{c\perp}$, $z' = 2z/a$ and $t' = t/t_0$, one may rewrite the *T*-state equation (9),

$$\text{Isotropic T-state equation in LTP} \begin{cases} \text{I}: \ -h\varphi' = \text{sgn}(e_\perp)/\tan(\psi-\varphi), \\ \text{II}: \ h' = \text{sgn}(e_\perp), \\ \text{III}: \ e_\perp \varphi' = h\dot{\varphi}, \\ \text{IV}: \ e_\perp' = \dot{h}. \end{cases} \quad (12)$$

Here, we omit the primes in $z'$ and $t'$. We start at the flux-free state. Let the applied magnetic field $h_{ax}$ along the $x$ axis, $h_{ax} = \dot{h}_{ax} t$. Taking into account the symmetry of the problem [$h(-z) = h(z)$, $\varphi(-z) = \varphi(z)$, $e(-z) = e(z)$ and $\psi(-z) = \psi(z) - \pi$], it suffices to solve the critical states in the region $0 \le z \le 1$. The boundary conditions are

$$\text{Boundary under } H_{ax} \begin{cases} \text{I: at } z=1, \ h = h_{ax}, \ \varphi = \varphi_a = 0, \\ \text{II:} \begin{cases} \text{at } z = z_p, \ h = 0, \ e_\perp = 0, \ \text{partial flux-penetration}, \\ \text{at } z = 0, \ e_\perp = 0, \qquad \text{full flux-penetration}. \end{cases} \end{cases} \quad (13)$$

Boundary I is required by the continuity of the magnetic field at the sample surface, $\mathbf{h} = \mathbf{h}_a$, assuming there is no surface barrier to vortex entry [12]. Note that the real boundary is $\mathbf{h} = \mathbf{h}_a + \mathbf{h}_s$ at the surface. $\mathbf{h}_s$ is the self-field at the surface reaching from outside and can be found by carrying out the Biot-Savart integration over the sample volume [44]. In the Bean model, the self-field at the surface from outside is $h_s = 1$. For the case $h_a \gg 1$, the self-field effect can be reasonably neglected, $\mathbf{h} = \mathbf{h}_a$. Boundary II relates to the continuity of the electric field at the flux front $z_p$ at which $h = 0$. Recalling that wherever $e_\perp \ne 0$ there occurs the current $J_{c\perp}$, the electric field $e_\perp$ has to vanish at $z = z_p$ which separates the flux-penetration region with $|J_\perp| = J_{c\perp}$ from the flux-free region with $J_\perp = 0$.

Increasing $h_{ax}$, the magnetic flux penetrates in from the surface $z = 1$. Once the sample is fully penetrated, the currents at the two sides of the symmetric axis $z = 0$ will touch each other. Since the symmetry gives $\psi(-z) = \psi(z) - \pi$, the azimuth of the current $\psi$ is discontinuous at $z = 0$. To avoid this discontinuous, $e_\perp$ must vanish at $z = 0$ such that there is no superconducting current.

Regarding the $T$-state equation (12) and the boundary condition (13), we have four equations for the four unknown quantities $h$, $e_\perp$, $\varphi$ and $\psi$, which are solvable without introducing the voltage-current law $E(J)$. In fact, the voltage-current dependence is not fully identified in terms of the unknown flux-flow resistivity. For instance, the flux-flow resistivity law contains the undetermined $\rho_f$, and the power law and the Bean model contains $E_0$. If $E_\perp$ and $J$ are obtained in the critical-state equations, then one can determine the complete voltage-current dependence. The solutions are obtained as

$$\text{Solution for isotropic LTP under } H_{ax} \begin{cases} h = z + h_{ax}(t) - 1, \ \varphi = 0, \ \psi = 0.5\pi, \\ e_\perp = \begin{cases} \dot{h}_{ax}(z + h_{ax} - 1), & \text{partial flux-penetration}, \\ \dot{h}_{ax} z, & \text{full flux-penetration}. \end{cases} \end{cases} \quad (14)$$

Under the sweep rate $\dot{h}_{ax}$ we find that $e_\perp > 0$, such that from the $T$-state equation II and Boundary I we obtain the solution of $h$. This leads to a penetration frontier $z_p = 1 - h_{ax}(t)$. Substituting $h$ in the $T$-state equation IV and then associating with the Boundary II, we have $e_\perp$. The $T$-state equation III has the general solution $\varphi = \Phi(z + \dot{h}_a t)$ [$\varphi = \Phi(0.5z^2 - z + \dot{h}_a tz)$ for full penetration case]. The arbitrary continuous differentiable function $\Phi$ is determined by the Boundary I, $\varphi(z=1) = 0$, which leads to $\varphi = 0$. It follows that $\psi = 0.5\pi$ from the $T$-state equation I. The solution for the full-penetration state remains Eq. (14) except $e_\perp = \dot{h}_{ax} z$. As far as the sample is not fully penetrated, the profiles $e(z)$ and $h(z)$ develop with the flux front moving inward. After the flux fully penetrates the slab, $e(z)$ is saturated with $e = \dot{h}_{ax} z$, but $h(z)$ continues to increase.

Now let us switching on $h_{ay}$ ($0 \leq \varphi_a \leq \pi$) while $h_{ay}$ keeps constant of $h_{ax}^0$, Fig. 1. The initial condition at the moment $t=0$ when we start to apply $h_{ay}$, is $h = h_{ax}^0$, $\varphi = 0$ and $\psi = \pi/2$. The boundary conditions are

$$\text{Boundary under } H_{ax} \text{ and } H_{ay} \begin{cases} \text{I: at } z=1,\ h = h_a = \sqrt{h_{ax}^2 + h_{ay}^2},\ \varphi = \arccos(h_{ay}/h_a), \\ \text{II:} \begin{cases} \text{at } z = z_p^+,\ h = 0,\ e_\perp = 0, & \text{partial flux-penetration}, \\ \text{at } z = 0,\ e_\perp = 0, & \text{full flux-penetration}. \end{cases} \end{cases} \quad (15)$$

Here, $z_p^+$ represents the penetration frontier in this case. Following the method for solution (14), we have

$$\text{Solution for isotropic LTP under } H_{ax} \text{ and } H_{ay} \begin{cases} h = z + h_a - 1,\ \varphi = \arctan(h_{ax}/h_y),\ \psi = \pi, \\ e_\perp = \begin{cases} \dot{h}_a(z + h_a - 1), & \text{partial flux-penetration}, \\ \dot{h}_a z, & \text{full flux-penetration}. \end{cases} \end{cases} \quad (16)$$

Here, $\dot{h}_a = h_a^{-1} \dot{h}_{ay} h_{ay}$. For the full flux-penetration, we have $e_\perp$ changed into $e_\perp = \dot{h}_a z$. In this case, the magnetic field profile $h(z)$ keeps a constant slope, while the electric field $e(z)$ has a varying gradient. $\mathbf{h}$ changes its magnitude and rotates during the penetration, while the direction of the current $\mathbf{j}$ is fixed at the $-x$ axis. Note that, although the direction of $\mathbf{j}$ is constant, its magnitude $j$ varies appropriately such that its projections $j_\parallel$ and $j_\perp$ on the varying field $\mathbf{h}$ fulfill the $T$-state equations and the critical-state restrictions.

## 2.3 Analyses of general transport problem

To trigger a general transport problem (GTP), we apply another constant field $\mathbf{H}_{az} = H_{az}\hat{\mathbf{z}}$. Through the conservation law $(H_z)' = 0$, the magnetic field $\mathbf{H}$ inside the superconducting slab consists of the in-plane $\mathbf{H}^\mathrm{P} = H_x \hat{\mathbf{x}} + H_y \hat{\mathbf{y}}$ and out-of-plane $\mathbf{H}_z = H_{az}\hat{\mathbf{z}}$. The superscript "P" tells that the quantity is in the slab plane. One thus finds

$$\mathbf{H} = H\hat{\mathbf{H}} = \mathbf{H}^\mathrm{P} + \mathbf{H}_z = H\hat{\mathbf{H}}^\mathrm{P} + H_{az}\hat{\mathbf{z}}, \quad (17)$$

where

$$\hat{\mathbf{H}} = \hat{\mathbf{H}}^\mathrm{P} \sin\theta + \hat{\mathbf{z}}\cos\theta = \hat{\mathbf{x}}\cos\varphi\sin\theta + \hat{\mathbf{y}}\sin\varphi\sin\theta + \hat{\mathbf{z}}\cos\theta. \quad (18)$$

Here, $\theta$ is the tilt angle between $\hat{\mathbf{H}}$ and $\hat{\mathbf{z}}$. Following the physical idea in [27], we can find such a set of solutions for the critical states in GTP: $\mathbf{E} = \mathbf{E}^\mathrm{P} + \mathbf{E}_z$, $\mathbf{H} = \mathbf{H}^\mathrm{P} + \mathbf{H}_z$ and $\mathbf{J} = \mathbf{J}^\mathrm{P} + \tilde{\mathbf{J}}$, in which $\mathbf{E}_z$ is a curl-free field expressed as the gradient of a scalar potential, $\mathbf{E}_z = -\nabla \Phi$. $\mathbf{E}^p$, $\mathbf{H}^p$ and $\mathbf{J}^p$ fulfill the Maxwell equations in LTP. Since $\mathbf{E}^p$ fulfills $\nabla \cdot \mathbf{E}^p = 0$, we have

$$\nabla \cdot \mathbf{E} = \nabla \cdot \mathbf{E}_z = -\nabla^2 \Phi = Q_{es}/\varepsilon_0. \tag{19}$$

which determines the scalar potential $\Phi$. Here, $Q_{es}$ is the charge density. Recalling the charge conservation law, the perturbation $\tilde{\mathbf{J}}$ of the current $\mathbf{J}$ in terms of $\mathbf{E}_z$ now relates to the generation of the electric charge, i.e. $\nabla \cdot \tilde{\mathbf{J}} = -\partial Q_{es}/\partial t$. These non-stationary currents can be neglected in the magneto-quasi-stationary regime.

We now express the critical current $\mathbf{J}_{c\perp}$ in the plane $\perp \mathbf{H}$ in terms of the in-plane $\mathbf{J}^p$ and $\mathbf{H}^p$. Shown in Fig. 3, the unit vector $\hat{\mathbf{l}}_i$ of the intersection line $\mathbf{l}_i$ between the plane $\mathbf{H} - \hat{\mathbf{z}}$ and the plane $\perp \mathbf{H}$ is

$$\hat{\mathbf{l}}_i = \hat{\mathbf{H}} \cos\theta - \hat{\mathbf{z}} \sin\theta = \hat{\mathbf{x}} \cos\theta \cos\varphi + \hat{\mathbf{y}} \cos\theta \sin\varphi - \hat{\mathbf{z}} \sin\theta. \tag{20}$$

**Figure 3.** Geometrical representation of the magnetic field and current density in isotropic LTP and GTP.

The projection of the current density $\mathbf{J}$ on the plane $\perp \mathbf{H}$ is $\mathbf{J}_\perp = |\mathbf{J}_\perp| \hat{\mathbf{J}}_\perp = \mathbf{J}_\perp^p + (\mathbf{J}_\parallel^p \cdot \hat{\mathbf{l}}_i) \hat{\mathbf{l}}_i = J_\perp^p \hat{\mathbf{H}}_\perp + J_\parallel^p \cos\theta \hat{\mathbf{l}}_i$. Combining this with Eq. (20) leads to

$$|\mathbf{J}_\perp| = J^p / \Omega(\psi, \varphi, \theta), \tag{21}$$

where we have introduced the notation

$$\Omega = [1 - \cos^2(\psi - \varphi) \sin^2\theta]^{-1/2}, \tag{22}$$

and the unit vector

$$\hat{\mathbf{J}}_\perp = \Omega[\cos(\psi - \varphi) \cos\theta \hat{\mathbf{l}}_i + \sin(\psi - \varphi) \hat{\mathbf{H}}_\perp^p] = \hat{J}_\perp^\parallel \hat{\mathbf{H}}^p + \hat{J}_\perp^\perp \hat{\mathbf{H}}_\perp^p + \hat{J}_\perp^z \hat{\mathbf{z}}, \tag{23}$$

where

$$\hat{J}_\perp^\parallel = \Omega \cos(\psi - \varphi) \cos^2\theta, \quad \hat{J}_\perp^\perp = \Omega \sin(\psi - \varphi), \quad \hat{J}_\perp^z = -\Omega \cos(\psi - \varphi) \cos\theta \sin\theta. \tag{24}$$

At the *T*-critical states, we obtain the critical-state condition for the in-plane current density $J^p$, $J_\parallel^p$ and $J_\perp^p$:

$$J^{\mathrm{p}} = J_{\mathrm{c}}^{\mathrm{p}} = \Omega J_{\mathrm{c}\perp}, \ J_{\perp}^{\mathrm{p}} = \Omega \sin(\psi - \varphi) J_{\mathrm{c}\perp}, \ J_{\parallel}^{\mathrm{p}} = \Omega \cos(\psi - \varphi) J_{\mathrm{c}\perp}. \tag{25}$$

In the $\hat{\mathbf{I}}_i - \hat{\mathbf{H}}_\perp^{\mathrm{p}}$ plane, the rotation matrix for $\hat{\mathbf{I}}_i = [1,0]^{\mathrm{T}}$ and $\hat{\mathbf{J}}_\perp^{\mathrm{p}} = [1 - \cos^2(\psi - \varphi)\sin^2\theta]^{-1/2}[\cos(\psi - \varphi)\cos\theta, \sin(\psi - \varphi)]^{\mathrm{T}}$ is

$$\mathbf{R}_{\mathrm{m}} = \begin{bmatrix} \cos\psi_\perp & -\sin\psi_\perp \\ \sin\psi_\perp & \cos\psi_\perp \end{bmatrix}, \tag{26}$$

which fulfills $\hat{\mathbf{J}}_\perp^{\mathrm{p}} = \mathbf{R}_{\mathrm{m}} \hat{\mathbf{I}}_p$. This enables one to determine the angle $\psi_\perp$ between $\hat{\mathbf{I}}_i$ and $\hat{\mathbf{J}}_\perp^{\mathrm{p}}$ as $\cos\psi_\perp = \Omega\cos(\psi - \varphi)\cos\theta$, $\sin\psi_\perp = \Omega\sin(\psi - \varphi)$ and

$$\tan\psi_\perp = \tan(\psi - \varphi)/\cos\theta. \tag{27}$$

This formula coincides with the appropriate one in [48]. The electric field $\mathbf{E}$ is determined by Eqs. (1) and (23), and the in-plane electric field $\mathbf{E}^{\mathrm{p}}$ is

$$\mathbf{E}^{\mathrm{p}} = E(\hat{J}_\perp^\parallel \hat{\mathbf{H}} + \hat{J}_\perp^\perp \hat{\mathbf{H}}_\perp) = E^{\mathrm{p}} \hat{\mathbf{E}}^{\mathrm{p}}, \tag{28}$$

where

$$E^{\mathrm{p}} = E[\hat{J}_\perp^\parallel \cos(\psi_E - \varphi) + \hat{J}_\perp^\perp \sin(\psi_E - \varphi)],$$

$$\hat{\mathbf{E}}^{\mathrm{p}} = (E/E^{\mathrm{p}})[(\hat{J}_\perp^\parallel \cos\varphi - \hat{J}_\perp^\perp \sin\varphi)\hat{\mathbf{x}} + (\hat{J}_\perp^\parallel \sin\varphi + \hat{J}_\perp^\perp \cos\varphi)\hat{\mathbf{y}}], \tag{29}$$

with its in-plane angle $\psi_E$ expressed as

$$\tan\psi_E = \frac{\hat{J}_\perp^\parallel \sin\varphi + \hat{J}_\perp^\perp \cos\varphi}{\hat{J}_\perp^\parallel \cos\varphi - \hat{J}_\perp^\perp \sin\varphi}. \tag{30}$$

Using Eq. (19), one finds an additional relation of $E$,

$$[-\Omega\cos(\psi - \varphi)\cos\theta\sin\theta E]' = Q_{es}. \tag{31}$$

In this case, $\mathbf{E}^{\mathrm{p}}$ is neither parallel nor perpendicular to $\mathbf{H}^{\mathrm{p}}$. Recalling that in LTP we also have $\mathbf{E}$ in an implicit direction at *CT*-critical states, then the *T*-critical state equations in GTP are obtained similarly with Eq. (11),

$$\text{Isotropic T-state equation in GTP} \begin{cases} \text{I: } -H^{\mathrm{p}}\varphi' = \Omega(\psi, \varphi, \theta)\cos(\psi - \varphi) J_{\mathrm{c}\perp}, \\ \text{II: } (H^{\mathrm{p}})' = \Omega(\psi, \varphi, \theta)\sin(\psi - \varphi) J_{\mathrm{c}\perp}, \\ \text{III: } (E^{\mathrm{p}})'\sin(\psi_E - \varphi) + \psi_E' E^{\mathrm{p}}\cos(\psi_E - \varphi) = \mu_0 \dot{H}^{\mathrm{p}}, \\ \text{IV: } (E^{\mathrm{p}})'\cos(\psi_E - \varphi) - \psi_E' E^{\mathrm{p}}\sin(\psi_E - \varphi) = -\mu_0 H^{\mathrm{p}}\dot{\varphi}. \end{cases} \tag{32}$$

There are six functions $H^{\mathrm{p}}$, $\psi$, $\varphi$, $\theta$, $\psi_E$ and $E^{\mathrm{p}}$ for six equations that are the *T*-state equations (32) and the additional relations (30) and (31).

The external magnetic fields $H_{ax}$ and $H_{ay}$ are applied successively as in LTP, and the boundary conditions in GTP remain Eqs. (13) and (15). Let us consider the special case where the applied field $H_{az}$ along the $z$ axis is much larger than the in-plane magnetic fields, $H_{az} \gg H_{ax}, H_{ay}$. We see immediately for this case that $\mathbf{H}$ is almost normal to the $xy$ plane, such that the inclined angle $\theta \approx 0$ [44]. Equations (22), (23), (25), (28) and (30) then tell us that $\Omega = 1$, $\mathbf{J}_\perp = |\mathbf{J}_\perp|\hat{\mathbf{J}}_\perp = \mathbf{J}^\text{p}$, $J^\text{p} = J^\text{p}_{c\perp}$, $\mathbf{E}^\text{p} = E^\text{p}\hat{\mathbf{J}}$ and $\psi_E = \psi$. The $T$-state equation (32) becomes

$$\text{Isotropic T-state equation in GTP} \begin{cases} \text{I: } -h\varphi' = \cos(\psi - \varphi), \\ \text{II: } h' = \sin(\psi - \varphi), \\ \text{III: } e'\sin(\psi - \varphi) + \psi'e\cos(\psi - \varphi) = \dot{h}, \\ \text{IV: } e'\cos(\psi - \varphi) - \psi'e\sin(\psi - \varphi) = -h\dot{\varphi}, \end{cases} \tag{33}$$

We have introduced the normalizations $h = 2H^\text{p}/(J_{c\perp}a)$, $e = 4E^\text{p}t_0/(\mu_0 J_{c\perp}a^2)$, $z = 2z/a$ and $t = t/t_0$.

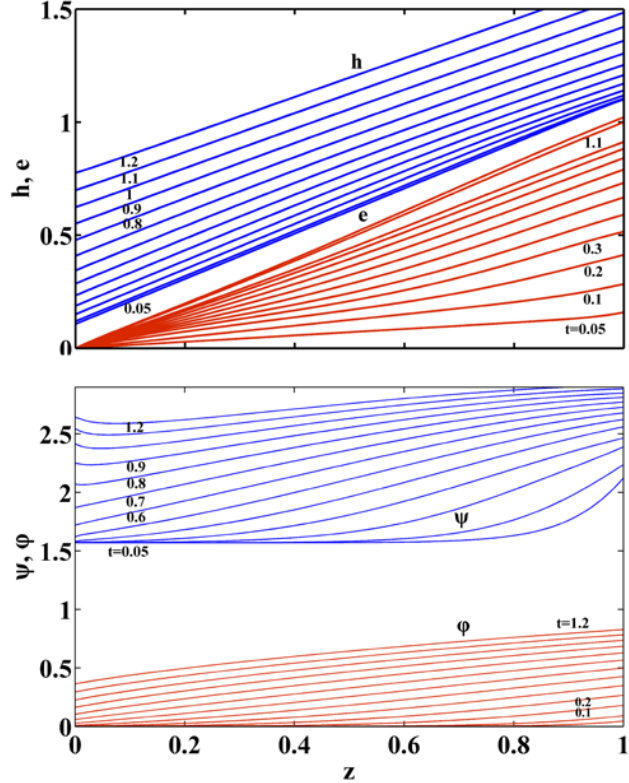

**Figure 4.** Profiles of the magnetic field $h(z)$, electric field $e(z)$, angle of the current $\psi$ and angle of the magnetic field $\varphi$ at the $T$-critical states of an isotropic slab in GTP. The applied magnetic fields $h^0_{ax} = 1.1$ and $h_{ay} = \dot{h}_{ay}t$ with $\dot{h}_{ay} = 1$ and $t = 0.05, 0.1, 0.2, 0.3, 0.4, 0.5, 0.6, 0.7, 0.8, 0.9, 1, 1.1$ and $1.2$. We start at the diamagnetic initial critical state described by Eq. (34) with $h^0_{ax} = 1.1$ and $\dot{h}_{ax} = 1$.

Let $H_{ax}$ be applied first. Recalling that in LTP we have the crossed magnetic field and current, $\psi - \varphi = 0.5\pi$ as in Eq. (14), then it is assumed this relation holds true here. Substituting this for Eq. (33) one obtains the simplified formulas: $-h\varphi' = 0$, $h' = 1$, $e' = \dot{h}$ and $\psi'e = h\dot{\varphi}$. One simply finds that the solution for this case coincides with the solution (14) in LTP.

For the subsequently switching on $H_{ay}$, one must solve the *T*-state equations (33) and the boundary condition (15). We consider the diamagnetic initial state which is obtained by first increasing $h_{ax}$ above the field of full flux penetration to $h_{ax}^0$ and then keeping $h_{ax}^0$ constant. At the moment $t = 0$, the beginning of applying $h_{ay}$, we have

$$\text{Initial condition (at t=0)} \begin{cases} \text{I: } h = h_{ax}^0 + z - 1, \; \varphi = 0, \\ \text{II: } e = \dot{h}_{ax} z \text{ and } \psi = 0.5\pi. \end{cases} \quad (34)$$

We consider the evolution of the critical states during an infinitesimal time period $[t, t + \Delta t]$, and then discretize the critical-state problem with respect to time. We denote the variables $h$, $\varphi$, $\psi$ and $e$ at the moment $t$ by $h_0$, $\varphi_0$, $\psi_0$ and $e_0$, and at the moment $t + \Delta t$ by $h_1$, $\varphi_1$, $\psi_1$ and $e_1$. One finds the discretized equations,

$\varphi_1' = -\cos(\psi_1 - \varphi_1)/h_1$, $\quad h_1' = \sin(\psi_1 - \varphi_1)$, $\quad \psi_1' = [(h_1 - h_0)\cos(\psi_1 - \varphi_1) + h(\varphi_1 - \varphi_0)\sin(\psi_1 - \varphi_1)]/e_1 \Delta t$ and $e_1' = [(h_1 - h_0)\sin(\psi_1 - \varphi_1) - h(\varphi_1 - \varphi_0)\cos(\psi_1 - \varphi_1)]/\Delta t$. If information at $t$ is known, then the remaining problem is to solve four one-order differential equations with four dependent variable $h_1$, $\varphi_1$, $\psi_1$ and $e_1$ as functions of the position $z$. Associated with the boundary condition (15), the problem to solve is a typical boundary value problem of ordinary differential equations, which apply to various numerical approaches. There is a good agreement between the present results and the results by Mikitik and Brandt [44]. We do not calculate the long-time evolution as in [27, 44] since we focus on the anisotropy effect on the critical states.

### 3. Anisotropic critical-state model

High-temperature oxide superconductors are roughly characterized by uniaxial anisotropy related to the oxygen vacancies; the axis *c* indicates the material anisotropy and the plane *ab* coincides with the superconducting CuO planes. Say, for example, $YBa_2Cu_3O_{7-y}$. Recent experiment [56] highlights that $YBa_2Cu_3O_{6.991}$ has an extra in-plane (plane *ab*) anisotropy due to the nanoscale clustering of oxygen vacancies along the Cu-O chains. Let the coordinate axes of $O-xyz$ aligned with the material principal axes *a*, *b* and *c* of an anisotropic biaxial type-II superconductor, Fig. 5. In the collective pinning theory [52], the pinning force $f_p$ per unit vortex length is obtained from the collective pinning energy $E_{pin}$ through $f_p(\theta, \varphi, \psi) \simeq f_p^c \xi(\theta, \varphi, \psi_p + 0.5\pi)/\xi_{ab}$. Here, $f_p^c$ is the pinning force of a vortex line aligned with the axis *c* and without anisotropy in the plane *ab*. $\xi(\theta, \varphi, \psi_p + 0.5\pi)$ relates to the radius of the vortex core $r_c(\theta, \varphi, \psi)$. $\xi_{ab}$ is the coherence length in the slab plane *ab*.

*3.1 Anisotropic pinning force and critical current density*

Based upon the physical concepts developed by Mikitik and Brandt [47, 48, 57], we rearrange $f_p(\theta, \varphi, \psi)$ in terms of the basis vectors $\hat{\mathbf{l}}_i$ and $\hat{\mathbf{l}}_{i,\perp}$, Fig. 5,

$$(f_{p,1}^2 + f_{p,2}^2)^2 = (f_p^c)^2 (\eta f_{p,1}^2 + 2\alpha f_{p,1} f_{p,2} + \beta f_{p,2}^2), \quad (35)$$

where

$$f_{p,1} = f_p \cos\psi_p, \quad f_{p,2} = f_p \sin\psi_p,$$

$$\eta(\varphi) = \zeta \cos^2\varphi + \zeta^{-1}\sin^2\varphi, \quad \alpha(\varphi,\theta) = -\sin\varphi\cos\varphi\cos\theta(\zeta - \zeta^{-1}),$$

$$\beta(\varphi,\theta) = \cos^2\theta(\zeta\sin^2\varphi + \zeta^{-1}\cos^2\varphi) + \varepsilon^2\sin^2\theta. \tag{36}$$

The characterizations of flux-pinning anisotropy are $\varepsilon = \lambda_{ab}/\lambda_c$, $\lambda_{ab} = \sqrt{\lambda_a\lambda_b}$, $\xi_{ab} = \sqrt{\xi_a\xi_b}$ and $\zeta = \lambda_a/\lambda_b$. $\lambda$ and $\xi$ means the GL penetration depth and coherence length in microscopic viewpoint. Subscripts $a$, $b$ and $ab$ means the quantities along axis $a$, $b$ and plane $ab$. The last term $\varepsilon^2\sin^2\theta$ in $\beta$ can be omitted under $\varepsilon^2\tan^2\theta \ll 1$. If $\theta = 0$, one finds from Eqs. (35) and (36) that $f_p = (f_p^c)\sqrt{\eta(\varphi+\psi_p)}$. For uniaxial superconductors, $\zeta = 1$, one finds $f_p = f_p^c\sqrt{\cos\psi_p^2 + \cos^2\theta\sin\psi_p^2}$. These results coincide with the appropriate ones in [57].

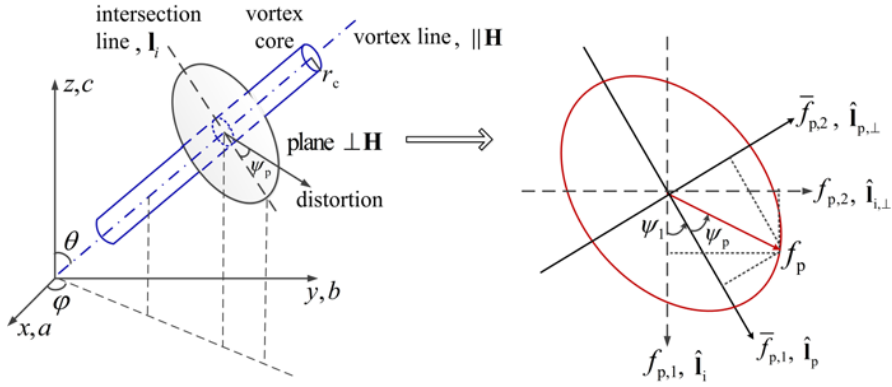

**Figure 5.** Left: An inclined vortex in an anisotropic superconductor. The axes $x$, $y$ and $z$ coincide with the material main axes $a$, $b$ and $c$ of the superconductor. Flux-line distortion occurs in the plane perpendicular to the vortex line; $\psi_p$ measures the angle between the distortion direction and the intersection line $\hat{\mathbf{l}}_i$ of this plane with the plane containing $z$ and the vortex line. $r_c$ is the radius of the vortex core. Right: Angular dependence of the anisotropic pinning force $f_p(\psi_p)$ in the plane perpendicular to the vortex line. $\psi_1$ measures the rotation from $\hat{\mathbf{l}}_i$ to the new principal axis $\hat{\mathbf{l}}_p$.

For $\alpha \neq 0$, we rotate the coordinates by an angle $\psi_1$ to eliminate the term with $f_{p,1}f_{p,2}$ in Eq. (35),

$$(\overline{f}_{p,1}^{\,2} + \overline{f}_{p,2}^{\,2})^2 = (f_p^c)^2(\overline{\eta}\,\overline{f}_{p,1}^{\,2} + \overline{\beta}\,\overline{f}_{p,2}^{\,2}), \tag{37}$$

where

$$\overline{f}_{p,1} = f_{p,1}\cos\psi_1 + f_{p,2}\sin\psi_1, \quad \overline{f}_{p,2} = -f_{p,1}\sin\psi_1 + f_{p,2}\cos\psi_1,$$

$$\overline{\eta} = 0.5[\eta + \beta + (\eta - \beta)\cos 2\psi_1] + \alpha\sin 2\psi_1,$$

$$\overline{\beta} = 0.5[\eta + \beta - (\eta - \beta)\cos 2\psi_1] - \alpha\sin 2\psi_1. \tag{38}$$

The rotation angle $\psi_1$ is determined by $2\alpha \cos 2\psi_1 = (\eta - \beta)\sin 2\psi_1$, such that

$$\psi_1 = \begin{cases} 0.5 \arctan[2\alpha/(\eta-\beta)], & \eta \neq \beta, \\ \pi/4, & \eta = \beta. \end{cases} \quad (39)$$

Equation (37) can be put into a quasi-elliptic form with respect to the angle $\psi_p$ in the plane $\perp \mathbf{H}$. Note that $\psi_p$ is now measured from the new principal axis $\hat{\mathbf{I}}_p$, Fig. 5. On the other hand, if $\alpha = 0$, no rotation should be made. We summary the anisotropic pinning force as:

$$f_p^2(\psi_p; \theta, \varphi) = f_{p0}^2 (\cos^2 \psi_p + \delta \sin^2 \psi_p), \quad (40)$$

where

$$f_{p0}^2 = \begin{cases} (f_p^c)^2 \bar{\eta}, & \alpha \neq 0, \\ (f_p^c)^2 \eta, & \alpha = 0, \end{cases} \text{ and } \delta = \begin{cases} \bar{\beta}/\bar{\eta}, & \alpha \neq 0, \\ \beta/\eta, & \alpha = 0. \end{cases} \quad (41)$$

Now, let us find the relationship between the anisotropic pinning force $f_p$ and the driving Lorentz force $f$. Note that $f$ acting on the vortex at an angle $\psi_f$ reaches its maximum value $f_\psi$ when its projection onto the direction $\psi_p$ of the pinning force balances against $f_p(\psi)$, i.e. $f_\psi \cos(\psi_f - \psi_p) = f_p(\psi_p)$, Fig. 6. The critical force $f_c$ at which the vortex starts to move is the minimum of $f_\psi$ over $\psi_p$. Note that, in some superconductors such as twin crystal, the condition that there is only one minimum may not occur.

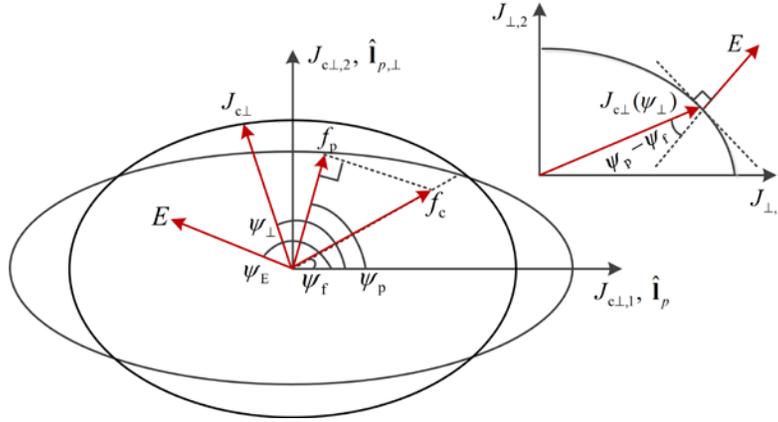

**Figure 6.** Schematic of the relationship between the pinning force $f_p$, the critical force $f_c$, the critical current density $J_{c\perp}(\psi_\perp)$ and the electric field $E(\psi_{E\perp})$. Insert: determination of the relation of $\mathbf{J}_{c\perp}(\psi_\perp)$ and $\mathbf{E}(\psi_{E\perp})$ in terms of the maximum projection rule.

Thus, the following equations [48]

$$\tan(\psi_f - \psi_p) = f_p'(\psi_p)/f_p(\psi_p), \quad f_c \cos(\psi_f - \psi_p) = f_p(\psi_p) \quad (42)$$

determine $f_c$ and $\psi_p$. Therefore, if $\delta \neq 0$, $\cos \psi_p \neq 0$ and $|\psi_f - \psi_p| < 0.5\pi$, one may use Eq. (42) to express them,

$$f_c(\psi_f) = f_{p0}/(\cos^2 \psi_f + \delta^{-1} \sin^2 \psi_f)^{1/2}, \quad \tan \psi_f = \delta \tan \psi_p. \quad (43)$$

At this point, anisotropy of the pinning force allows the direction of the flux-line velocity deviating from the direction of the driving Lorentz force in the plane $\perp \mathbf{H}$, $\psi_f \neq \psi_p$. Recalling that the critical force $f_c$ relates to the critical current density $\overline{\mathbf{J}}_{c\perp}$ through $\overline{\psi}_\perp = \psi_f + \pi/2$ and $J_{c\perp}(\psi_\perp)\Phi_0 = f_c(\psi_f)$ where $\overline{\psi}_\perp$ is the angle of $\overline{\mathbf{J}}_{c\perp}$ in the plane $\perp \overline{\mathbf{H}}$, Fig. 6, one can then determine $J_{c\perp}(\psi_\perp)$. On the other hand, if $\delta = 0$ one simply finds $\tan\psi_f = 0$, $f_c = f_{p0}$, $\psi_\perp = \pi/2$ and $J_{c\perp}(\psi_\perp) = f_{p0}/\Phi_0$. Thus, we obtain $\overline{J}_{c\perp}(\overline{\psi}_\perp)$ as

$$J_{c\perp}(\psi_\perp) = \begin{cases} f_{p0}\Phi_0^{-1}(\sin^2\psi_\perp + \delta^{-1}\cos^2\psi_\perp)^{-1/2}, & \delta \neq 0, \\ f_{p0}\Phi_0^{-1}, & \delta = 0. \end{cases} \quad (44)$$

Measurements [58] on $YBa_2Cu_3O_{7-y}$ thin films confirm the angular and field dependence $J_{c\perp}(\theta, H)$, $J_{c\perp}$ variation as a function of the tilt angle $\theta$ of magnetic field with respect to material axis $c$. A general dependence of $J_{c\perp}(\psi, \varphi, \theta)$ is given in Eq. (44), which can be reduced to $J_{c\perp}(\theta, H)$ under some particular cases.

*3.2 Material conductivity law*

The direction of the electric field $\mathbf{E}$ is described by its angle $\psi_{E\perp} = \psi_p + \pi/2$ in the plane $\perp \mathbf{H}$ according to $\mathbf{E} = \mathbf{B} \times \mathbf{v}_v$. We have $\tan(\psi_\perp - \psi_{E\perp}) = \tan(\psi_f - \psi_p) = f_p'(\psi_p)/f_p(\psi_p) = \psi_f'(\psi_p)f_c'(\psi_f)/f_c(\psi_f) - [\psi_f'(\psi_p) - 1]\tan[\psi_f(\psi_p) - \psi_p]$ from Eq. (42), and thus

$$\tan\psi_{E-\perp} = -\tan(\psi_\perp - \psi_{E\perp}) = -f_c'(\psi_f)/f_c(\psi_f) = -J_{c\perp}'(\psi_\perp)/\overline{J}_{c\perp}(\psi_\perp), \quad (45)$$

where $\psi_{E-\perp}$ signifies the direction of $\mathbf{E}$ with respect to $\mathbf{J}_{c\perp}$ in the plane $\perp \mathbf{H}$. This expression coincides with Mikitik and Brandt's proposal [41] obtained from the creep activation barrier. Note that Eq. (45) is valid independent with the specific form of pinning force $f_p(\psi; \theta, \varphi)$.

We finds the consistency between Eq. (45) and the maximal projection rule [14, 32, 33],

$$\partial\mathbf{J}_{c\perp} \perp \mathbf{E} \Leftrightarrow \max(\mathbf{J}_{c\perp} \cdot \hat{\mathbf{E}}) \Leftrightarrow \tan\psi_{E-\perp} = -J_{c\perp}'(\psi_\perp)/J_{c\perp}(\psi_\perp). \quad (46)$$

For the magnitude of the electric field $\mathbf{E}$, we obtain the resistivity law for flux transport in anisotropic superconductor, $J_\perp \cos(\psi_E - \psi_\perp) = \rho_\perp E_\perp$. In the isotropic case, we neglect anisotropy of the flux-flow resistivity $\rho_\perp$. We may remind the readers that even in a uniaxial superconductor, the flux-flow resistivity shows an a angular dependence with $\theta$ as well as anisotropy in the plane perpendicular to the vortex [59].

*3.3 General anisotropic critical-state model*

The $O$ zone in GDCSM becomes a cylinder in the frame of $J_{\perp,1} - J_{\perp,2} - J_\|$, since $J_c(J_{\perp,1}, J_{\perp,2})$ sketches a circle. If anisotropic flux-line pinning is introduced, $J_c(J_{\perp,1}, J_{\perp,2})$ then changes into an ellipse, Eq. (44). Thus, one may obtain a cylindroid for anisotropic cases in $J_{\perp,1} - J_{\perp,2} - J_\|$, Fig. 7. The material conductivity law at the critical states in isotropic superconductors is generalized into

$$\mathbf{J}_\perp = \begin{cases} J_{c\perp}\hat{\mathbf{E}}_\perp, & \text{isotropy} \\ J_{c\perp}^a \hat{\mathbf{E}}_\perp^a, & \text{anisotropy} \end{cases}, \quad E_\perp \neq 0,$$

$$\mathbf{J}_\| = J_{c\|}\hat{\mathbf{E}}_\|, \qquad E_\| \neq 0. \tag{47}$$

In anisotropic superconductors the direction of $\mathbf{J}_\perp$ deviates from $\mathbf{E}_\perp$ so, $\mathbf{J}_\perp$ is parallel to $\hat{\mathbf{E}}_\perp^a$ given by Eq. (52) and $J_{c\perp}^a$ is described with Eq. (44).

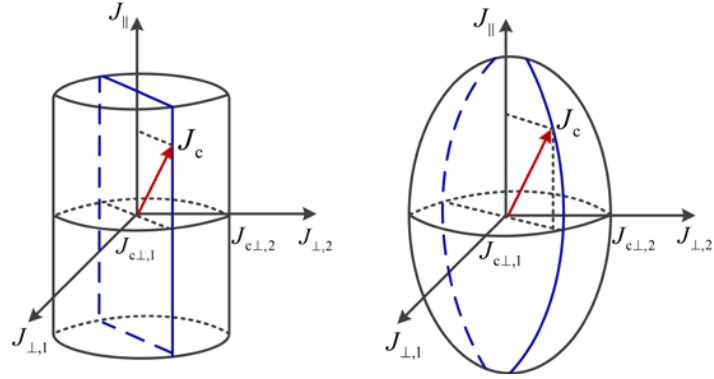

**Figure 7.** Anisotropic critical-state models in coordinates $J_{\perp,1} - J_{\perp,2} - J_\|$. Left: Extension from GDCSM. Right: Extension from BM critical-state model or the extended elliptical critical model.

The elliptical critical model describes the critical current density $\mathbf{J}_{c\perp}$ as an ellipse in the $J_\| - J_\perp$ plane, viz., $\sin^2\phi/J_\perp^2 + \cos^2\phi/J_\|^2 = 1/J_c^2$, where $\phi$ is the angle between $\mathbf{J}_{c\perp}$ and the magnetic field $\mathbf{H}$, Fig. 2. Inside the ellipse, the so-called O zone, no flux depinning and cutting occur. The flux transport and cutting simultaneously occur everywhere at and outside the ellipse, except at the axes $J_\perp = 0$ (no transport) and $J_\| = 0$ (no cutting). $J_c$ along the axis $J_\perp = 0$ (or $J_\| = 0$) represents the threshold for depinning (or cutting), denoted by $J_{cd} = J_c|\sin\phi|$ (or $J_{cc} = J_c|\cos\phi|$). Assuming that the ellipse model in the $J_\| - J_\perp$ plane holds true while its principal axis $\bar{J}_\perp$ changes with $\psi_\perp$ in the plane $J_{\perp,1} - J_{\perp,2}$ due to anisotropic flux-line pinning [see Eq. (44)], thereby the ellipse is extended to a quasi-ellipsoid in the coordinates $J_\| - J_{\perp,1} - J_{\perp,2}$, Fig. 7:

$$\frac{\sin^2\phi}{J_{c\perp}^2(\psi_\perp)} + \frac{\cos^2\phi}{J_{c\|}^2} = \frac{1}{J_c^2(\psi,\phi)}. \tag{48}$$

If the superconductor is isotropic, $J_{c\perp}(\psi_\perp)$ then degrades to a constant $J_{c\perp}$, and the quasi-ellipsoid becomes a standard one. The material law at the critical states is the same as Eq. (47). Alternatively, one may introduce the relations for the magnitudes of $\mathbf{J}$ and $\mathbf{E}$, viz., $E_\perp = \rho_\perp J_\perp$ and $E_\| = \rho_\| J_\|$ where $\rho_\perp$ and $\rho_\|$ are the nonlinear effective resistivities [12].

*3.4 Analyses of general transport problem*

*3.4.1 Anisotropic T-state equations*

Let us first analyze LTP (longitudinal transport problem) in an anisotropic slab. Rotating the intersection line $\hat{\mathbf{I}}_i$ by the angle $\psi_1$, we obtain the principal axis $J_{\perp,1}$ of the $J_{c\perp}(\psi_\perp)$ ellipse. In LTP, $\theta = \pi/2$, the intersection line is parallel with $z$ axis, Fig. 3. The current $\mathbf{J}_\perp$ thus equals to the in-plane current component $\mathbf{J}_\perp$, and $\psi_\perp$ has a simple relation $\psi_\perp = 0.5\pi - \psi_1$. Also, one finds $\alpha = \beta = 0$ from Eq. (36) such that the principal axis coincides with $\hat{\mathbf{I}}_i$, i.e. $\psi_1 = 0$. According to Eq. (40), $J_{c\perp}(\psi_\perp) = f_{p0}/\Phi_0$ describes a circle in the plane $\perp \mathbf{H}$. Thus, we have $J_{c\perp}'(\psi_\perp) = 0$, and from Eq. (45) it follows that $\psi_{E-\perp} = 0$. Finally, we obtain $\mathbf{E} = \text{sgn}(J_\perp)E\hat{\mathbf{H}}_\perp = E_\perp \hat{\mathbf{H}}_\perp$, which is the same as in the isotropic case, Eq. (5). The anisotropic LTP is therefore the same as the isotropic one, except for the expression of $J_{c\perp}$. The anisotropic $J_{c\perp}(\psi_\perp) = f_{p0}/\Phi_0$ in which $f_{p0} = f_p^c \eta^{0.5}$ and $\eta(\varphi) = \zeta \cos^2\varphi + \zeta^{-1}\sin^2\varphi$, Eqs. (36) and (41). While the isotropic case with $\zeta = 1$ gives $f_{p0} = f_p^c$. In Section 4, we will apply the anisotropic LTP to analyze the critical states in an anisotropic sample exposed to an in-plane rotating magnetic field.

Now we turn to general transport problem, GTP. The inclined magnetic field $\mathbf{H}$ leads to the plane $\perp \mathbf{H}$ out of the $x-y$ plane such that the projection of $\mathbf{J}$ on the plane $\perp \mathbf{H}$ is $\mathbf{J}_\perp$ instead of the in-plane $\mathbf{J}_\perp^p$, Fig. 3; the anisotropic pinning further causes a deviation $\psi_{E-\perp}$ of the direction of $\mathbf{E}$ from $\mathbf{J}_\perp$ in the plane $\perp \mathbf{H}$. Now the direction of the electric field $\mathbf{E}$ reads as

$$\hat{\mathbf{E}} = \cos\psi_{E-\perp}\hat{\mathbf{J}}_\perp + \sin\psi_{E-\perp}(\hat{\mathbf{H}} \times \hat{\mathbf{J}}_\perp). \tag{49}$$

The current density $\mathbf{J}_\perp$ in the plane $\perp \mathbf{H}$ relates to the in-plane quantities $\mathbf{H}^p$ and $\mathbf{J}^p$ through the same expressions (21)-(25) as in the isotropic case except for $J_{c\perp}(\psi_\perp)$. $\psi_\perp$ now measures the angle between $\mathbf{J}_\perp$ and the principal axis $\hat{\mathbf{I}}_p$ due to the coordinates rotation $\psi_1$, Fig. 5.

The principal axis in this plane is $\hat{\mathbf{I}}_p = \cos\psi_1 \hat{\mathbf{I}}_i + \sin\psi_1 \hat{\mathbf{H}}_\perp^p$. We then have the rotation matrix for the two vectors $\hat{\mathbf{I}}_p = [\cos\psi_1, \sin\psi_1]^T$ and $\hat{\mathbf{J}}_\perp = [1-\cos^2(\psi-\varphi)\sin^2\theta]^{-1/2}[\cos(\psi-\varphi)\cos\theta, \sin(\psi-\varphi)]^T$:

$$\mathbf{R}_m = \begin{bmatrix} \cos\psi_\perp & -\sin\psi_\perp \\ \sin\psi_\perp & \cos\psi_\perp \end{bmatrix}, \tag{50}$$

which fulfills $\hat{\mathbf{J}}_\perp = \mathbf{R}_m \hat{\mathbf{I}}_p$, such that $\cos(\psi_\perp + \psi_1) = [1-\cos^2(\psi-\varphi)\sin^2\theta]^{-1/2}\cos(\psi-\varphi)\cos\theta$, $\sin(\psi_\perp + \psi_1) = [1-\cos^2(\psi-\varphi)\sin^2\theta]^{-1/2}\sin(\psi-\varphi)$ and

$$\tan(\psi_\perp + \psi_1) = \tan(\psi-\varphi)/\cos\theta, \ \cos\theta \neq 0. \tag{51}$$

The deviation angle $\psi_1$ is already given in Eqs. (36) and (39). Recalling that for isotropic GTP the critical current $J_c^p$

depends on $\psi - \varphi$, $\theta$ and $B$, the additional dependence of $J_c^p$ on $\varphi$ occurs for anisotropic situation in terms of $J_c^p = J_c^p[\psi - \varphi, \theta, J_{c\perp}^p(B, \psi_\perp)]$ and $\psi_\perp = \psi_\perp[\psi - \varphi, \theta, \psi_1(\theta, \varphi)]$. Substituting Eqs. (18) and into Eq. (49), one obtains $\hat{\mathbf{H}} \times \hat{\mathbf{J}}_\perp = \Omega[-\cos\theta \sin(\psi - \varphi)\hat{\mathbf{H}}^p + \cos(\psi - \varphi)\cos\theta \hat{\mathbf{H}}_\perp^p + \sin\theta \sin(\psi - \varphi)\hat{\mathbf{z}}]$ and

$$\hat{\mathbf{E}} = \hat{E}_\| \hat{\mathbf{H}}^p + \hat{E}_\perp \hat{\mathbf{H}}_\perp^p + \hat{E}_z \hat{\mathbf{z}}, \tag{52}$$

where

$$\hat{E}_\| = \Omega[\cos\psi_{E-\perp} \cos(\psi - \varphi)\cos^2\theta - \sin\psi_{E-\perp} \sin(\psi - \varphi)\cos\theta],$$

$$\hat{E}_\perp = \Omega[\cos\psi_{E-\perp} \sin(\psi - \varphi) + \sin\psi_{E-\perp} \cos(\psi - \varphi)\cos\theta],$$

$$\hat{E}_z = -\Omega[\cos\psi_{E-\perp} \cos(\psi - \varphi)\cos\theta\sin\theta - \sin\psi_{E-\perp} \sin(\psi - \varphi)\sin\theta]. \tag{53}$$

Now the electric field $\mathbf{E}$ in the coordinates $O - xyz$ is expressed as $\mathbf{E} = E[(\hat{E}_\| \cos\varphi - \hat{E}_\perp \sin\varphi)\hat{\mathbf{x}} + (\hat{E}_\| \sin\varphi + \hat{E}_\perp \cos\varphi)\hat{\mathbf{y}} + \hat{E}_z \hat{\mathbf{z}}]$ in terms of Eq. (53). If $\psi_E$ signifies the direction of the in-plane $\mathbf{E}^p$, $\hat{\mathbf{E}}^p = [\cos\psi_E, \sin\psi_E, 0]$, the following equations are obtained:

$$\tan\psi_E = \frac{\hat{E}_\| \sin\varphi + \hat{E}_\perp \cos\varphi}{\hat{E}_\| \cos\varphi - \hat{E}_\perp \sin\varphi}. \tag{54}$$

The $T$-state equations for anisotropic superconductors have the same form with Eq. (32) in the isotropic case, however, $J_{c\perp}$ and $\psi_E$ are with different expressions, Eqs. (44) and (54). In fact, the anisotropic pinning entering in the critical states of anisotropic superconductors is characterized by the projection angle $\psi_E$ of the inclined electric field $\mathbf{E}$ (49) and the critical-state restrictions $J_{c\|}^p$ and $J_{c\perp}^p$ (25). $\psi_E$ relates to the anisotropy through the deviation angle $\psi_{E-\perp}$ (45), while $J_\|^p$ (or $J_\perp^p$) is dependent with $J_{c\perp}(\psi_\perp)$. $\psi_{E-\perp}$ is also a function of $J_{c\perp}(\psi_\perp)$, thus we find from Eqs. (39), (44) and (51) that the critical state problem corrected for anisotropy stems from the anisotropic superconducting parameters ($\xi_a$, $\xi_b$, $\lambda_a$, $\lambda_b$ and $\lambda_c$) and the azimuth angles $\theta$, $\varphi$ and $\psi$ of $\mathbf{H}$ and $\mathbf{J}^p$. For simplicity we first consider a uniaxial anisotropic sample ($\zeta = 1$). In this case, one finds from Eqs. (36) that $\eta(\varphi) = 1$, $\alpha(\varphi, \theta) = 0$ and $\beta(\varphi, \theta) = \cos^2\theta$. No rotation of the coordinates is needed, $\psi_1 = 0$, such that $f_{p0} = f_p^c$ and $\delta = \cos^2\theta$ in terms of Eqs. (41).

If $\theta = \pi/2$ as for LTP, we find from Eqs. (21), (25) and (44) that $\hat{\mathbf{J}}_\perp = \hat{\mathbf{H}}_\perp^p \sin(\psi - \varphi)/|\sin(\psi - \varphi)|$ and $J_\perp^p = J_{c\perp}(\psi_\perp)\sin(\psi - \varphi)/|\sin(\psi - \varphi)|$. Since $\delta = 0$ and $f_{p0} = f_p^c$ we have $J_{c\perp} = J_c^c = f_p^c/\Phi_0$ from Eq. (44). As for the electric field, Eqs. (45) and (49) tell that $\psi_{E-\perp} = 0$ and $\mathbf{E} = E^p\hat{\mathbf{E}} = E^p\hat{\mathbf{J}}_\perp$. Combining these results one recovers the isotropic $T$-critical state in LTP, Eq. (9).

If $\theta = 0$, we have $\hat{\mathbf{J}}_\perp = \cos(\psi - \varphi)\hat{\mathbf{H}}^p + \sin(\psi - \varphi)\hat{\mathbf{H}}_\perp^p$, $J_\perp^p = \sin(\psi - \varphi)J_{c\perp}(\psi_\perp)$, $J_\|^p = \cos(\psi - \varphi)J_{c\perp}(\psi_\perp)$, $\psi_\perp = \psi - \varphi$, $J_{c\perp} = J_c^c$, $\tan\psi_{E-\perp} = 0$ and $\psi_E = \psi$. These results imply that the current $\mathbf{J}^p$ is parallel with the electric field $\mathbf{E}^p$, and the anisotropy has nothing to do with the critical states. In other words, the critical states in this case are consistent with the isotropic *T*-critical states in GTP, Eq. (32).

*3.4.2 Calculations*

Now let us cope with a more complex situation, in which the external magnetic fields always fulfill $H_{az} \gg H_{ax}, H_{ay}$ acting on an anisotropic biaxial superconductors, $\theta = 0$ and $\zeta \neq 1$. We stress that an anisotropic critical-state problem with a different set of boundary conditions has already been treated by Badía and López [60]. From Eqs. (36) one finds $\eta = \zeta \cos^2 \varphi + \zeta^{-1} \sin^2 \varphi$, $\alpha = -\sin\varphi\cos\varphi(\zeta - \zeta^{-1})$ and $\beta = \zeta \sin^2 \varphi + \zeta^{-1} \cos^2 \varphi$ such that $\psi_1 = -\varphi$ in Eq. (39). Also we have $\bar{\eta} = \zeta$, $\bar{\beta} = \zeta^{-1}$, $f_{p0}^2 = (f_p^c)^2 \zeta$ and $\delta = \zeta^{-2}$ from Eqs. (38) and (41). The angle of the critical current density $\mathbf{J}_{c\perp}$ is $\psi_\perp = \psi$, Eq. (51). Then, according to Eqs. (44), (45), (53) and (54), one obtains $J_c^p$, $\psi_{E-\perp}$ and $\psi_E$. We are now ready to set down the anisotropic *T*-state equation in GTP,

$$\text{Anisotropic T-state equation in GTP} \begin{cases} \text{I: } -H^p \varphi' = \cos(\psi - \varphi) J_c^p, \\ \text{II: } (H^p)' = \sin(\psi - \varphi) J_c^p, \\ \text{III: } (E^p)' \sin(\psi_E - \varphi) + \psi_E' E^p \cos(\psi_E - \varphi) = \mu_0 \dot{H}^p, \\ \text{IV: } (E^p)' \cos(\psi_E - \varphi) - \psi_E' E^p \sin(\psi_E - \varphi) = -\mu_0 H^p \dot{\varphi}, \end{cases} \quad (55)$$

where

$$J_c^p(\psi, \varphi) = J_c^c (\zeta^{-1} \sin^2 \psi + \zeta \cos^2 \psi)^{-1/2},$$

$$\psi_E(\psi) = \psi + \psi_{E-\perp}, \quad \psi_{E-\perp}(\psi) = \arctan\left(\frac{\zeta^{-1} - \zeta}{\zeta^{-1} \tan\psi + \zeta \tan^{-1}\psi}\right). \quad (56)$$

Note that in this case, the critical current density coincides with its projection on the slab plane, $J_c = J_c^p$.

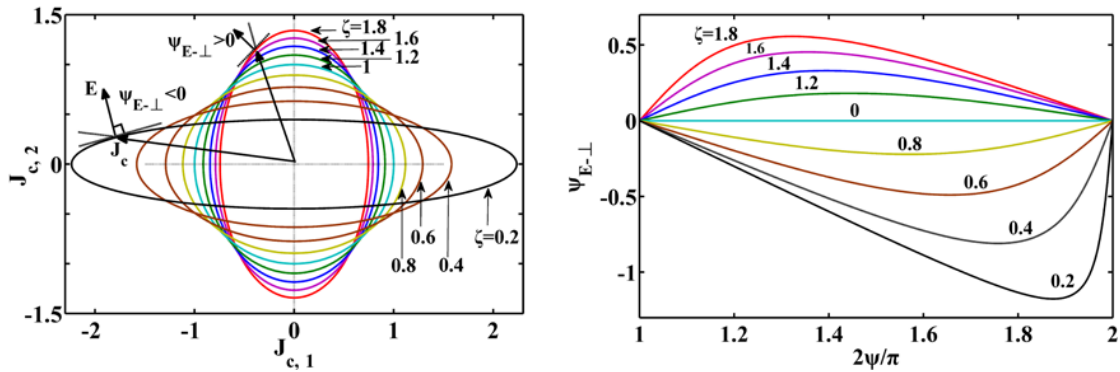

**Figure 8.** The angular dependences of $J_c$ (in units $J_c^c$) and the deviation angle $\psi_{E-\perp}$ in an anisotropic biaxial slab. The external excitations are the same as in Fig. 4 and $\zeta = 0.2, 0.4, 0.6, 0.8, 1, 1.2, 1.4, 1.6$ and $1.8$.

From Fig. 8 we observe that, the ellipse $J_c(\psi)$ takes the axis $J_{c,1}$ as the long axis, and the intercept at the long axis decreases as $\zeta$ changes from 0.2 to 1. When crossing over $\zeta = 1$ and increasing $\zeta$, one obtains the ellipse which has the increasing long axis located on the $J_{c,2}$ axis. In fact, $\zeta<1$ corresponds to $\delta>1$, and the expression (56) of $J_c$ describes the appropriate ellipse in Fig. 8. $\psi_\perp=\psi$ tells that the principal axis $J_{c,1}$ coincides with the axis $x$ as well as the material axis $a$. In physical sense, an anisotropic superconductor which has a smaller penetration parameter $\lambda_a$ along the material axis $a$, i.e. $\zeta=\lambda_a/\lambda_b<1$, corresponds to a $J_c(\psi)$ ellipse characterized by the long axis $a$; if $\lambda_a$ exceeds $\lambda_b$, then the $J_c(\psi)$ ellipse takes the long axis $b$. We also find that an enhanced in-plane anisotropy (larger $|\zeta-1|$) gives rives to a lager deviation angle $\psi_{E-\perp}$. In the interval of interest, $\pi/2<\psi<\pi$, $\psi_{E-\perp}$ with $\zeta<1$ always corresponds to a negative value, while $\psi_{E-\perp}>0$ in the case of $\zeta>1$. This can be explained by applying the maximum projection rule on the $J_c(\psi)$ ellipse, Fig. 8.

The anisotropic $T$-state equation (55) differs from the isotropic one (33) regarding $J_c$ and $\psi_E$. When the superconductor is of uniaxial anisotropy, $\zeta=1$, $J_c$ and $\psi_E$ are reduced to $J_c^c$ and $\psi$. The anisotropic $T$-state equations (55) are thus reduced to Eq. (33). Introducing the normalizations $h=2H^p/J_c^c a$, $e=4E^p t_0/\mu_0 J_c^c a^2$, $z=2z/a$ and $t=t/t_0$, we have

$$\text{Anisotropic T-state equation in GTP} \begin{cases} \text{I:} & -h\varphi'=\vartheta(\varphi,\psi)\cos(\psi-\varphi), \\ \text{II:} & h'=\vartheta(\varphi,\psi)\sin(\psi-\varphi), \\ \text{III:} & e'\sin(\psi_E-\varphi)+\psi_E' E\cos(\psi_E-\varphi)=\dot{h}, \\ \text{IV:} & e'\cos(\psi_E-\varphi)-\psi_E' e\sin(\psi_E-\varphi)=-h\dot{\varphi}. \end{cases} \tag{57}$$

where $\vartheta(\varphi,\psi)=(\zeta^{-1}\sin^2\psi+\zeta\cos^2\psi)^{-1/2}$, and $\psi_E$ is given by Eq. (56). First, we consider only $H_{ax}$ acting on the anisotropic superconductor, thus we simply obtain such critical states,

$$\text{Solution for anisotropic GTP under } H_{ax} \begin{cases} h=\zeta^{1/2}z+\dot{h}_{ax}t-\zeta^{1/2}, \ \varphi=0, \ \psi_E=\psi=0.5\pi, \\ e=\begin{cases} \dot{h}_{ax}[z-\zeta^{-1/2}(\zeta^{1/2}-\dot{h}_{ax}t)], & \text{partial flux-penetration}, \\ \dot{h}_{ax}z, & \text{full flux-penetration}. \end{cases} \end{cases} \tag{58}$$

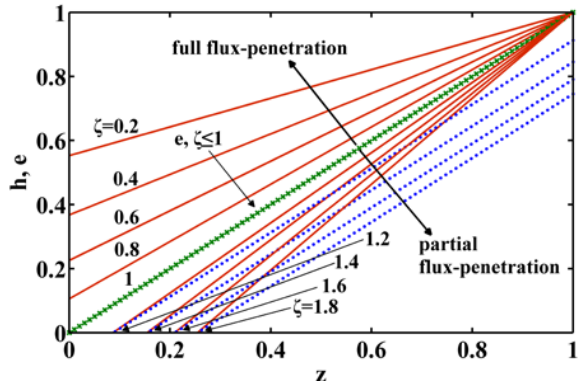

**Figure 9.** Diamagnetic initial state profiles $h(z)$ (solid lines) and $e(z)$ (dotted lines) in an anisotropic biaxial slab. The applied field is characterized by $\dot{h}_{ax}=1$ and $h_{ax}^0=1$. The in-plane anisotropy parameter $\zeta$ =0.2, 0.4, 0.6, 0.8, 1, 1.2, 1.4, 1.6 and 1.8. The profile of $h(z)$ at $\zeta=1$ and the profiles of $e(z)$ at $\zeta\leq 1$ coincide with each other (cross).

Increasing $h_{ax}$ to $h_{ax}^0$ forms the diamagnetic initial state for the subsequent imposition of $H_{ay}$ on the anisotropic slab. For $\zeta < 1$, the slab is fully penetrated with magnetic flux, and the electric field $e$ is saturated at $e = \dot{h}_{ax} z$, Fig. 9. While for $\zeta > 1$, the flux partially penetrates in, such that $e = 0$ at the flux front which changes with time. We consider the fixed boundary problem for computation convenience so, in the following we calculate the cases of $\zeta < 1$.

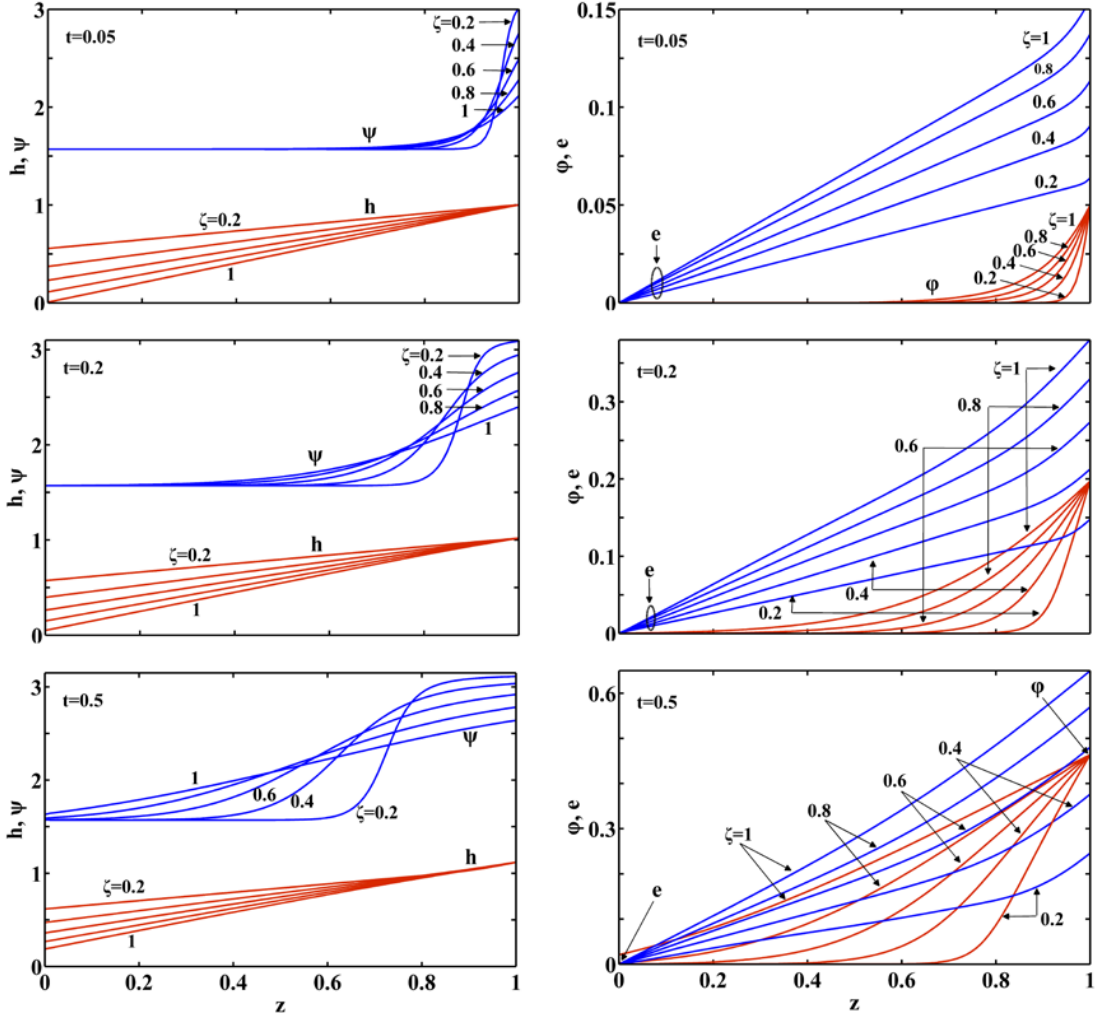

**Figure 10.** The critical-state profiles in an anisotropic biaxial slab at $\zeta$ =0.2, 0.4, 0.6, 0.8 and 1. The applied magnetic fields $h_{ax}^0 = 1$ and $h_{ay} = \dot{h}_{ay} t$ with $\dot{h}_{ay} = 1$. We start at the diamagnetic initial critical state described by Eq. (58) with $h_{ax}^0 = 1$ and $\dot{h}_{ax} = 1$.

At the moment $t = 0$ corresponding to the beginning of switching on $H_{ay}$, the initial condition is given by Eq. (58) and the boundary is given by Eq. (15). From Eq. (56) we obtain

$$\psi_E' = [1 - \frac{(\zeta^{-1} - \zeta)(\zeta^{-1} \cos^{-2} \psi - \zeta \sin^{-2} \psi)}{(\zeta^{-1} \tan \psi + \zeta \tan^{-1} \psi)^2 + (\zeta^{-1} - \zeta)^2}] \psi' = \varpi(\psi)\psi'. \tag{59}$$

Then one may rewrite the *T*-state equation (57) in favor of numerical procedure,

$$\text{Anisotropic T-state equation in GTP} \begin{cases} \text{I: } \varphi' = -\vartheta \cos(\psi - \varphi)/h, \\ \text{II: } h' = \vartheta \sin(\psi - \varphi), \\ \text{III: } \psi' = \varpi^{-1} e^{-1} [\dot{h} \cos(\psi_E - \varphi) + h\dot{\varphi} \sin(\psi_E - \varphi)], \\ \text{IV: } e' = \dot{h} \sin(\psi_E - \varphi) - h\dot{\varphi} \cos(\psi_E - \varphi). \end{cases} \quad (60)$$

After discretization with time, Eq. (60) is a set of first-order ordinary differential equations with the dependent variables $\varphi$, $h$, $\psi$ and $e$. Employing the numerical methods as in the isotropic GTP, we obtain the critical states at $t = 0.05$ shown in Fig. 10. An enhanced in-plane anisotropy, i.e. smaller $\zeta$, always moderates the gradient of the magnetic field $h$ and the electric field $e$ along the thickness $z$. Since for different $\zeta$ the boundary condition requires $h = h_a$ at $z = 1$, the reduction in the gradient of $h$ at an enhanced in-plane anisotropy raises the profile of $h$. In contrast, the profile of the electric field $e$ is suppressed regarding the decreasing gradient and the boundary condition $e = 0$ at $z = 1$. From the profiles of $\varphi(z)$, one finds an increasing slope with respect to $z$ at a stronger in-plane anisotropy, thus a lower distribution curve of $\varphi$.

The profiles of $\psi(z)$ at $t = 0.05$ shows roughly a knot at $z_k \approx 0.9$, near the slab surface. At $z > z_k$, an enhanced in-plane anisotropy causes an increasing slope and thus a raised profile curve; while at $z < z_k$, we observe a smaller slope of the profile curve accompanied by a lower profile curve. The different characterizations of the critical state profiles reflect the nonlinear effect of the in-plane anisotropy. Compared to those at $t = 0.05$, the further evolution ($t = 0.2$ and $0.5$) of the critical states exhibits a more significant effect of the anisotropy on the critical states. The dependences of $h$, $\varphi$, $\psi$ and $e$ with the in-plane anisotropy $\zeta$ are qualitatively consistent with those in the case of $t = 0.05$. Note that the knot $z_k$ tends to disperse as it shifts inward during the critical-state evolution.

## 4. Other example: three-zone structure in field cooled rotation experiment

Now we consider a superconducting infinite slab subjected to an external magnetic field $\mathbf{H}_a(t) = H_a \hat{\mathbf{H}}_a(t)$ with constant magnitude $H_a$ but changing direction $\hat{\mathbf{H}}_a = \hat{\mathbf{x}} \cos \varphi_a + \hat{\mathbf{y}} \sin \varphi_a$. Here, $\varphi_a = \dot{\varphi}_a t$, and $\dot{\varphi}_a$ is a constant angular velocity. It is assumed that at $t = 0$ a uniform magnetic field equal to the applied field exists in the superconducting slab, $\mathbf{H}(z, 0) = H_a \hat{\mathbf{x}}$, and thus a nonmagnetic initial state is formed. The evolution of the field profile $\mathbf{H}(z, t)$ occurs in the slab, and the magnetic response is evaluated by the magnetic moment, $\mathbf{M} \equiv \langle \mathbf{H}(z) \rangle - \mathbf{H}_a$, which can be measured experimentally. This regime corresponds to the situation in the field cooled rotation experiment [61, 62].

Recalling the analyses in Section 3.4.1, we know that the anisotropic flux-pinning enters into the critical states by only changing the value of $J_{c\perp}$, Fig. 11,

$$J_{c\perp} = f_{p0}/\Phi_0 = f_p^c \eta^{0.5}/\Phi_0 = J_c^c (\zeta \cos^2 \varphi + \zeta^{-1} \sin^2 \varphi)^{0.5}. \quad (61)$$

We want to stress that such anisotropic critical states have been investigated by Badía and López [60], who apply the phenomenological elliptic model and pseudoisotropic model to analyze the magnetic response of a slab in the field cooled rotation experiment. Since we do not consider the anisotropy effect on the flux cutting, the critical state equations and results are corrected slightly compared to Badía-López model. The phenomenological anisotropic parameter and the angle between the magnetic field and a given axis in Badía-López model now have definite physical essence. They refer to the anisotropic parameter $\zeta = \lambda_a / \lambda_b$ which is ratio of the London penetration depths along the two in-plane material principal axes $a$ and $b$, and the angle $\varphi$ between the magnetic field and axis $a$.

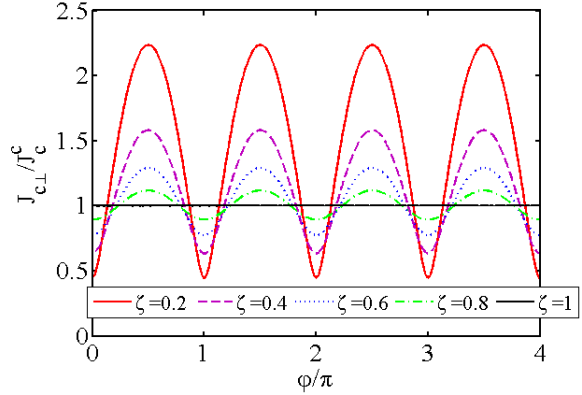

**Figure 11.** $J_{c\perp}$ variation with magnetic field rotation $\varphi$ in an anisotropic biaxial slab in a longitudinal rotating field.

At the threshold for the onset of flux-line cutting, we have $-h\varphi' = \mathrm{sgn}(e_\parallel)\chi$ [see the first equation in the *CT*-state equations (11)], where $\chi = J_{c\parallel}/J_{c\perp}$. This can be reformulated as $-\varphi' = \mathrm{sgn}(e_\parallel)\chi h^{-1}$. Following the idea by Clem and Perez-Gonzalez [21], one may take $\varphi'$ to be constants independent with magnetic induction, such that $-\varphi' = \mathrm{sgn}(e_\parallel)\chi h_a^{-1}$. At the beginning of the magnetic field rotation, the profiles of critical states develop a three-zone structure, Fig. 12. Over the *V*-shaped profile $h(z)$, a minimum occurs at $z = z_v$. The rotation $\varphi$ of the magnetic field within the slab obey $h_a\varphi' = \chi$ throughout the slab. At surface $z = 1$ we have $\varphi(z=1) = \varphi_a$, such that the magnetic field rotation stops at $z_c = 1 - \chi^{-1}h_a\varphi_a$, i.e. $\varphi(z = z_c) = 0$. A *O* zone ($0 \leq z < z_c$) is the region from the middle plane $z = 0$ to $z = z_c$, in which $h = h_a$, $\varphi = 0$, and neither flux-line transport nor cutting occurs. The region from $z = z_c$ to the *V*-shape minimum $z = z_v$ is a $C_-$ zone ($z_c < z < z_v$), where the vortex is at the threshold of flux-line cutting. While, a $C\_T_+$ state occurs in the region from the minimum $z = z_v$ to the surface $z = 1$. In the $C\_T_+$ zone ($z_v \leq z \leq 1$), both flux-line transport and cutting occur.

In the $C_-$ zone ($z_c \leq z \leq z_v$), the solutions of *C*-state equations (10) subject to the boundary conditions $h(z = z_c) = h_a$, $\varphi(z = z_c) = 0$ and $e_\parallel(z = z_c) = 0$ are

$$h = h_a \cos[\chi h_a^{-1}(z - z_c)],\ \varphi = \chi h_a^{-1}(z - z_c),$$
$$e_\parallel = -\chi^{-1}\dot{\varphi}_a h_a^2 \sin[\chi h_a^{-1}(z - z_c)],\ j_\perp = -\chi \sin[\chi h_a^{-1}(z - z_c)]. \tag{62}$$

The $C_-$ zone always holds true if $\chi < 1$ such that $|j_\perp| = J_\perp/J_{c\perp} < 1$.

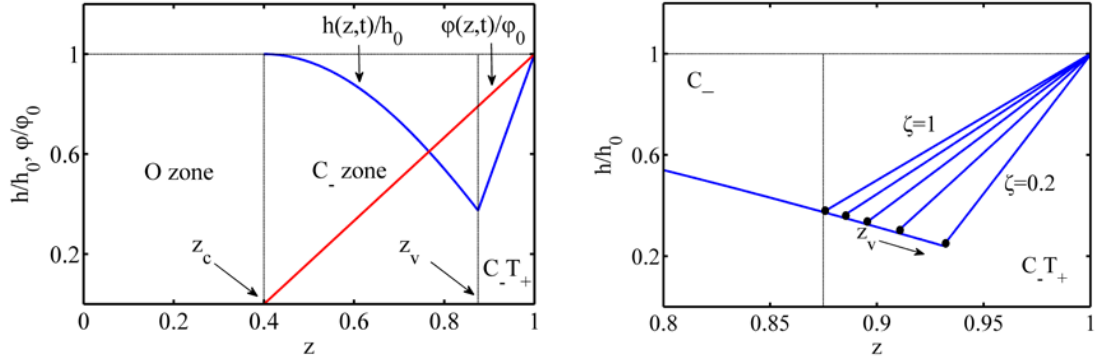

**Figure 12.** Left: Three-zone structure [$O$/$C_-$/$C_-T_+$] developed in an isotropic superconducting slab ($\zeta = 1$) subjected to a rotating magnetic field $\mathbf{H}_a(t) = H_a \hat{\mathbf{H}}_a(t)$ with constant $H_a$ and changing $\hat{\mathbf{H}}_a = \hat{\mathbf{x}}\cos\varphi_a + \hat{\mathbf{y}}\sin\varphi_a$. We start at the nonmagnetic initial state $\mathbf{h}(z,0) = h_a \hat{\mathbf{x}}$. For this plot, the time $t = 1.5$, applied field $h_a = 0.2$, ratio $J_{c\|}/J_{c\perp} = \chi = 0.5$ and applied angular velocity $\dot{\varphi}_a = 1$. Right: Field profiles $h(x,t)$ in the $C_-T_+$ zone for the anisotropic parameters $\zeta = 0.2, 0.4, 0.6, 0.8$ and $1$.

In the $C_-T_+$ zone ($z_v < z \leq 1$), the solutions of $CT$-state equations (11) subject to the boundary conditions $h(z=1) = h_a$, $\varphi(z=1) = \varphi_a$, $h(z=z_v) = h_v$, $e_\|(z=z_v) = -\chi^{-1}\dot{\varphi}_a h_a^2 \sin[\chi h_a^{-1}(z_v - z_c)]$ and $e_\perp(z=z_v) = 0$ are

$$h = z + h_a - 1, \quad \varphi = \chi h_a^{-1}(z - z_c),$$

$$e_\perp = \chi^{-1} h_a \dot{\varphi}_a (z + h_a - 1) - \chi^{-1} h_a^2 \dot{\varphi}_a \cos[\chi h_a^{-1}(z - z_c)] - \chi^{-2} h_a^2 \dot{\varphi}_a \sin[\chi h_a^{-1}(z - z_v)],$$

$$e_\| = -\chi^{-2} h_a^2 \dot{\varphi}_a - \chi^{-1} h_a^2 \dot{\varphi}_a \sin[\chi h_a^{-1}(z - z_c)] + \chi^{-2} h_a^2 \dot{\varphi}_a \cos[\chi h_a^{-1}(z - z_v)]. \tag{63}$$

Continuity of $h(z,t)$ at $z = z_v$ yields $h_a \cos[\chi h_a^{-1}(z_v - z_c)] = z_v + h_a - 1$, which determines the value of $z_v$. This relation can be rewritten as $h_a \cos\varphi_v = h_v$, where $h_v = z_v + h_a - 1$ and $\varphi_v = -\chi h_a^{-1}(z_v - z_c)$. The above equations remain valid if the magnetic field $h_v$ at the minimum of $V$-shape profile $z_v$ is not less than 0. This requires $\varphi_a = \chi h_a^{-1}(1 - z_c) \leq 0.5\pi + \chi$. On the other hand, to form the three-zone structure one must have $z_c \geq 0$ which leads to $\varphi_a < \chi h_a^{-1}$. This condition should be fulfilled when $h_v$ reduces to zero, and in this sense one may write $h_a < \chi(0.5\pi + \chi)^{-1}$ which determines the effective range of $h_a$.

As $\varphi_a$ increases from zero to $(0.5\pi + \chi)$, the magnetic flux front $z_c$ of the $C_-/C_-T_+$ zone penetrates in, and the intersection point $z_v$ of the two profiles $h(z,t)$ enters more deeply into the superconductor, Fig. 13. When $\varphi_a = 0.5\pi + \chi$, $h_v$ is reduced to zero and a quasisteady-state distribution of $h(x,t)$ is achieved. The minimum $z_v$ of the $V$-shape profile locates at $z_v = z_0$ [Fig. 13, curves $t = 0.5(\pi + 1)$]. As $\varphi_a$ proceeds to increase, $\varphi_a > 0.5\pi + \chi$, the profiles of $h(x,t)$ throughout the region $0 \leq z \leq 1$ remain unchanged. However, $\varphi(x,t)$ in the $C_-T_+$ zone ($z_0 \leq z \leq 1$) follows the variation $\varphi = \chi h_a^{-1}(z - z_c)$ under the variation of $\varphi_a$ [Fig. 14, curves $t > 0.5(\pi + 1)$].

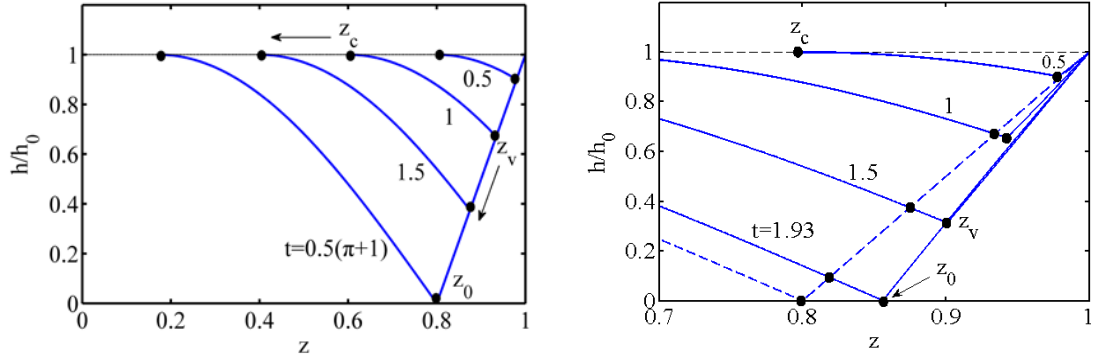

**Figure 13.** Evolution of $h(x,t)$ in isotropic (Left) and anisotropic (Right, $\zeta = 0.5$ and the dash lines refer to those in isotropic case) slab. Sample parameters and applied magnetic field are as in Fig. 12.

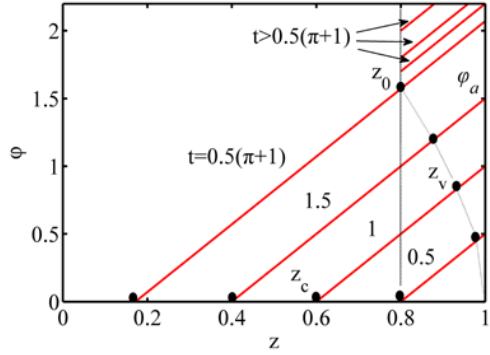

**Figure 14.** Evolution of $\varphi(x,t)$. Sample parameters and applied magnetic field are as in Fig. 12.

Incorporating anisotropic flux-line pinning, the solutions for $C_-$ zone remain unchanged, but $j_\perp$ now must satisfy $|j_\perp| < j_{c\perp}$ where $j_{c\perp} = (\zeta \cos^2 \varphi + \zeta^{-1} \sin^2 \varphi)^{0.5}$. Notice that $j_{c\perp}$ is a periodic function with a period of $\pi$, the minimum of $j_{c\perp}(\varphi)$ is $\zeta^{0.5}$ and the maximum is $\zeta^{-0.5}$, Fig. 11 (as illustrated in Section 3.4.2, we only consider the case of $\zeta < 1$). So, we have the condition for validation of the $C_-$ state, $\chi < \zeta^{-1}$. The second formula in the *CT*-state equations (11) changes into $h' = (\zeta \cos^2 \varphi + \zeta^{-1} \sin^2 \varphi)^{0.5}$, in which $\varphi$ is the same with the one in the isotropic case. Using numerical approach one may find the value of $h(x,t)$ under different anisotropic parameters $\zeta$, Fig. 12. The profile $h(x,t)$ shifts toward the surface $z=1$ with an enhanced in-plane anisotropy (smaller $\zeta$), and the region $C_-T_+$ shrinks. As the applied field rotates initially, for an applied rotation $z_v$ shifts toward the surface $z=1$ compared to those in isotropic case, and the slope of $h(x,t)$ is no longer constant, Fig. 13. At an earlier time the decoupling point $z_0$ is defined. As rotation proceeds, $h(x,t)$ keeps no longer stationary. This is because, the changing $\varphi$ in the region $z_0 \leq z \leq 1$ requires the appropriate changing $h(x,t)$ to fulfill the *T*-state condition $h' = (\zeta \cos^2 \varphi + \zeta^{-1} \sin^2 \varphi)^{0.5}$. Thus, a complex structure with more critical states arises, and the three-zone structure may develop into four (or more)-zone structure.

The magnetic profiles $h(z,t)$ for $\varphi_a \geq 0.5\pi + \chi$ reflect the observations in field cooled experiments [62]. The experimental sample is $YBa_2Cu_3O_7$ which shows evidently an anisotropic pinning. Experiments show a so-called magnetic flux bifurcation ($z_0$), which separates the flux lines into two groups, one of which fixed to the sample (region $0 \leq z \leq z_0$), while the other rotates frictionally relative to the sample (region $z_0 \leq z \leq 1$).

The magnetic flux bifurcation occurs if the applied field amplitude $h_a$ is within a low-field range $h_a < \chi(0.5\pi + \chi)^{-1}$. When $h_a > (0.5\pi\chi^{-1} + 1)^{-1}$, the three-zone structure and magnetic flux bifurcation are no longer sustained. In this case, as rotation increases, the vortex fluxes penetrate in and reach the midplane $z = 0$ of the slab. After that, the magnetic flux bifurcation vanishes, and an evolutionary successive and smooth profile of $h(z,t)$ may exist throughout a half of the slab thickness $0 \leq z \leq 1$.

## 5. Conclusions

In this article, we develop a critical-state model accounting for anisotropic flux-line pinning in a type-II superconducting slab. The anisotropic critical current density and its relation with the electric field are formulated upon the collective pinning theory. The critical-state equations are thus corrected for the anisotropic pinning. Using the critical-state equations and the appropriate boundary conditions, we calculate the longitudinal and general transport problem in the slab with and without anisotropy. The importance of pinning anisotropy is emphasized in the evolutionary critical-state profiles. We apply the critical-state model in a practical example, viz. the magnetic response in a slab rotating in a uniform in-plane magnetic field. A three-zone structure including *O*-, *C*- and *CT*- zone is predicted. The isotropic critical model gives a stationary profile $h(x,t)$ after the magnetic flux bifurcation is achieved. While in the anisotropic case, the profile $h(x,t)$ becomes unsteady, and a four (or more) zone structure is developed.

A restriction of the proposed model is that only the anisotropic pinning is considered; anisotropy surely arises in flux-cutting mechanism, and the underlying physics should be developed, which is not limited to a phenomenological description. On the other hand, the analytical approach based on the Maxwell equations and $\mathbf{E} - \mathbf{J}$ characteristic, as in the presented model, may encounter computation difficulties in the problem with long-time loading and arbitrarily oriented field. The variational statements [14, 34, 36] are outstanding in solving arbitrary complex critical-state problem.

**Acknowledgments**

This work was performed with supports from the National Natural Science Foundation of China (11372120, 11032006 and 11421062), National Key Project of Magneto-Restriction Fusion Energy Development Program (2013GB110002) and Fundamental Research Funds for the Central Universities (lzujbky-2014-227).